\documentclass[twocolumn]{aastex63}

\usepackage{xspace}
\newcommand{\Kepler}{\emph{Kepler}\xspace}
\newcommand{\TESS}{\emph{TESS}\xspace}
\newcommand{\rhocirc}{\rho_{\rm circ}}
\newcommand{\thisstar}{TOI-216\xspace}
\newcommand{\thisstarinn}{TOI-216b\xspace}
\newcommand{\thisstarout}{TOI-216c\xspace}
\newcommand{\inn}{b\xspace}
\newcommand{\out}{c\xspace}
\shorttitle{\thisstar}
\shortauthors{Dawson et al.}
\usepackage{multirow}
\received{July 31, 2020}
\revised{December 12, 2020}
\accepted{January 4, 2021}
\submitjournal{AJ}

\begin{document}

\title{Precise transit and radial-velocity characterization of a resonant pair: a warm Jupiter TOI-216c and eccentric warm Neptune TOI-216b}
\correspondingauthor{Rebekah I. Dawson}
\email{rdawson@psu.edu}

\author[0000-0001-9677-1296]{Rebekah I. Dawson}
\affiliation{Department of Astronomy \& Astrophysics, Center for Exoplanets and Habitable Worlds, The Pennsylvania State University, University Park, PA 16802, USA}
\author[0000-0003-0918-7484]{Chelsea~ X.~Huang}
\affiliation{Department of Physics and Kavli Institute for Astrophysics and Space Research, Massachusetts Institute of Technology, Cambridge, MA 02139, USA}
\affiliation{Juan Carlos Torres Fellow}
\author[0000-0002-9158-7315]{Rafael Brahm}
\affiliation{Facultad de Ingeniería y Ciencias, Universidad Adolfo Ib\'a\~nez, Av.\ Diagonal las Torres 2640, Pe\~nalol\'en, Santiago, Chile}
\affiliation{Millennium Institute for Astrophysics, Chile}
\author[0000-0001-6588-9574]{Karen A.\ Collins}
\affiliation{Harvard-Smithsonian Center for Astrophysics, 60 Garden St., Cambridge, MA 02138, USA}
\author[0000-0002-5945-7975]{Melissa J. Hobson} 
\affiliation{Millennium Institute for Astrophysics, Chile}
\affiliation{Instituto de Astrof\'isica, Facultad de F\'isica, Pontificia Universidad Cat\'olica de Chile}
\author[0000-0002-5389-3944]{Andr\'es Jord\'an} 
\affiliation{Facultad de Ingeniería y Ciencias, Universidad Adolfo Ib\'a\~nez, Av.\ Diagonal las Torres 2640, Pe\~nalol\'en, Santiago, Chile}
\affiliation{Millennium Institute for Astrophysics, Chile}
\author[0000-0002-3610-6953]{Jiayin Dong}  
\affiliation{Department of Astronomy \& Astrophysics, Center for Exoplanets and Habitable Worlds, The Pennsylvania State University, University Park, PA 16802, USA}
\affiliation{Center for Computational Astrophysics, Flatiron Institute, 162 Fifth Avenue, New York, NY 10010, USA}
\author[0000-0002-0076-6239]{Judith Korth}  
\affiliation{Rheinisches Institut f\"ur Umweltforschung, Abteilung Planetenforschung an der Universit\"at zu K\"oln, Universit\"at zu K\"oln, Aachenerstraße 209, 50931 K\"oln, Germany}
\author{Trifon Trifonov} 
\affiliation{Max Planck Institute for Astronomy, Koenigstuhl 17, D-69117 Heidelberg, Germany}
\author{Lyu Abe} 
\affiliation{Université Côte d'Azur, Observatoire de la Côte d'Azur, CNRS, Laboratoire Lagrange Bd de l'Observatoire, CS 34229, 06304 Nice cedex 4, France
 } %
 \author{Abdelkrim Agabi}  
\affiliation{Université Côte d'Azur, Observatoire de la Côte d'Azur, CNRS, Laboratoire Lagrange Bd de l'Observatoire, CS 34229, 06304 Nice cedex 4, France
 } 
\author[0000-0002-1560-4590]{Ivan Bruni}  
\affiliation{INAF OAS, Osservatorio di Astrofisica e Scienza dello Spazio di Bologna via Piero Gobetti 93/3, Bologna}
\author{R. Paul Butler} 
\affiliation{Carnegie Institution for Science, Earth \& Planets Laboratory, 5241 Broad Branch Road NW, Washington DC 20015, USA}
\author{Mauro Barbieri} 
\affiliation{INCT, Universidad de Atacama, calle Copayapu 485, Copiap\'o, Atacama, Chile}
\author[0000-0003-2781-3207]{Kevin I.\ Collins} 
\affiliation{George Mason University, 4400 University Drive, Fairfax, VA, 22030 USA}
\author[0000-0003-2239-0567]{Dennis M.\ Conti}  
\affiliation{American Association of Variable Star Observers, 49 Bay State Road, Cambridge, MA 02138, USA}
\author{Jeffrey D. Crane} 
\affiliation{Observatories of the Carnegie Institution for Science, 813 Santa Barbara Street, Pasadena, CA 91101}
\author[0000-0001-7866-8738]{Nicolas Crouzet}
\affiliation{European Space Agency (ESA), European Space Research and Technology Centre (ESTEC), Keplerlaan 1, 2201 AZ Noordwijk, The Netherlands} 
\author[0000-0002-3937-630X]{Georgina Dransfield} 
\affiliation{University of Birmingham, School of Physics \& Astronomy, Edgbaston, Birmingham B15 2TT, United Kingdom}
\author[0000-0002-5674-2404]{Phil Evans}
\affiliation{El Sauce Observatory, Coquimbo Province, Chile}
\author[0000-0001-9513-1449]{N\'{e}stor Espinoza} 
\affiliation{Space Telescope Science Institute, 3700 San Martin Drive, Baltimore, MD 21218, USA}
\author[0000-0002-4503-9705]{Tianjun~Gan}
\affil{Department of Astronomy, Tsinghua University, Beijing 100084, China}
\author[0000-0002-7188-8428]{Tristan Guillot} 
\affiliation{Université Côte d'Azur, Observatoire de la Côte d'Azur, CNRS, Laboratoire Lagrange Bd de l'Observatoire, CS 34229, 06304 Nice cedex 4, France
 }
\author{Thomas Henning} 
\affiliation{Max Planck Institute for Astronomy, Koenigstuhl 17, D-69117 Heidelberg, Germany}
\author[0000-0001-6513-1659]{Jack J. Lissauer}
\affiliation{NASA Ames Research Center, Moffett Field, CA 94035}
\author[0000-0002-4625-7333]{Eric L. N. Jensen}  
\affiliation{Dept.\ of Physics \& Astronomy, Swarthmore College, Swarthmore PA 19081, USA}
\author{Wenceslas Marie Sainte} 
\affiliation{Institut Paul Emile Victor, Concordia Station, Antarctica}
 \author{Djamel M\'{e}karnia} %
\affiliation{Université Côte d'Azur, Observatoire de la Côte d'Azur, CNRS, Laboratoire Lagrange Bd de l'Observatoire, CS 34229, 06304 Nice cedex 4, France
 } 
\author[0000-0002-9810-0506]{Gordon Myers}  
\affiliation{AAVSO, 5 Inverness Way, Hillsborough, CA 94010, USA}
\author{Sangeetha Nandakumar} 
\affiliation{INCT, Universidad de Atacama, calle Copayapu 485, Copiap\'o, Atacama, Chile}
\author{Howard M. Relles} 
\affiliation{Harvard-Smithsonian Center for Astrophysics, 60 Garden St., Cambridge, MA 02138, USA}
\author{Paula Sarkis} 
\affiliation{Max Planck Institute for Astronomy, Koenigstuhl 17, D-69117 Heidelberg, Germany}
\author{Pascal Torres} 
\affiliation{Instituto de Astrof\'isica, Facultad de F\'isica, Pontificia Universidad Cat\'olica de Chile}
\affiliation{Millennium Institute for Astrophysics, Chile}
\author{Stephen Shectman} 
\affiliation{Observatories of the Carnegie Institution for Science, 813 Santa Barbara Street, Pasadena, CA 91101}
 \author[0000-0003-3914-3546]{Fran\c{c}ois-Xavier Schmider}
\affiliation{Université Côte d'Azur, Observatoire de la Côte d'Azur, CNRS, Laboratoire Lagrange Bd de l'Observatoire, CS 34229, 06304 Nice cedex 4, France
 } 
\author[0000-0002-1836-3120]{Avi Shporer} 
\affiliation{Department of Physics and Kavli Institute for Astrophysics and Space Research, Massachusetts Institute of Technology, Cambridge, MA 02139, USA}
\author[0000-0003-2163-1437]{Chris Stockdale}  
\affiliation{Hazelwood Observatory, Australia}
\author{Johanna Teske} 
\affiliation{Observatories of the Carnegie Institution for Science, 813 Santa Barbara Street, Pasadena, CA 91101}
\affiliation{NASA Hubble Fellow}
\author[0000-0002-5510-8751]{Amaury H.M.J. Triaud} 
\affiliation{University of Birmingham, School of Physics \& Astronomy, Edgbaston, Birmingham B15 2TT, United Kingdom}
\author[0000-0002-6937-9034]{Sharon Xuesong Wang} 
\affiliation{Observatories of the Carnegie Institution for Science, 813 Santa Barbara Street, Pasadena, CA 91101}
\author{Carl Ziegler} 
\affil{Dunlap Institute for Astronomy and Astrophysics, University of Toronto, 50 St. George Street, Toronto, Ontario M5S 3H4, Canada}
\author{G. Ricker}  
\affiliation{Department of Physics and Kavli Institute for Astrophysics and Space Research, Massachusetts Institute of Technology, Cambridge, MA 02139, USA}
\author[0000-0001-6763-6562]{R. Vanderspek}  
\affiliation{Department of Physics and Kavli Institute for Astrophysics and Space Research, Massachusetts Institute of Technology, Cambridge, MA 02139, USA}
\author{David W. Latham} 
\affiliation{Harvard-Smithsonian Center for Astrophysics, 60 Garden St., Cambridge, MA 02138, USA}
\author[0000-0002-6892-6948]{S. Seager}  
\affiliation{Department of Physics and Kavli Institute for Astrophysics and Space Research, Massachusetts Institute of Technology, Cambridge, MA 02139, USA}
\affiliation{Department of Earth, Atmospheric, and Planetary Sciences, Massachusetts Institute of Technology, Cambridge, MA 02139, USA}
\affiliation{Department of Aeronautics and Astronautics, Massachusetts Institute of Technology, Cambridge, MA 02139, USA}
\author[0000-0002-4265-047X]{J. Winn}
\affiliation{Department of Astrophysical Sciences, Princeton University, 4 Ivy Lane, Princeton, NJ 08540, USA}
\author[0000-0002-4715-9460]{Jon M. Jenkins} 
\affiliation{NASA Ames Research Center, Moffett Field, CA 94035}
\author[0000-0002-0514-5538]{L. G. Bouma}  
\affiliation{Department of Astrophysical Sciences, Princeton University, 4 Ivy Lane, Princeton, NJ 08540, USA}
\author[0000-0002-0040-6815]{Jennifer~A.~Burt}  
\affiliation{Jet Propulsion Laboratory, California Institute of Technology, 4800 Oak Grove drive, Pasadena CA 91109, USA}
\author[0000-0002-9003-484X]{David~Charbonneau}  
\affiliation{Harvard-Smithsonian Center for Astrophysics, 60 Garden St., Cambridge, MA 02138, USA}
\author{Alan~M.~Levine}
\affiliation{Department of Physics and Kavli Institute for Astrophysics and Space Research, Massachusetts Institute of Technology, Cambridge, MA 02139, USA}
\author{Scott~McDermott} 
\affiliation{Proto-Logic LLC, 1718 Euclid Street NW, Washington, DC 20009, USA} 
\author[0000-0002-8058-643X]{Brian McLean}
\affiliation{Mikulski Archive for Space Telescopes, Space Telescope Science Institute, 3700 San Martin Drive, Baltimore, MD 21218, USA}
\author[0000-0003-4724-745X]{Mark E. Rose} 
\affiliation{NASA Ames Research Center, Moffett Field, CA 94035}
\author[0000-0001-7246-5438]{Andrew~Vanderburg}
\affiliation{Department of Astronomy, The University of Texas at Austin, Austin, TX 78712, USA}
\affiliation{NASA Sagan Fellow}
\author[0000-0002-5402-9613]{Bill Wohler} 
\affiliation{SETI Institute, Mountain View, CA 94043, USA}
\affiliation{NASA Ames Research Center, Moffett Field, CA 94035}

\begin{abstract}
TOI-216 hosts a pair of warm, large exoplanets discovered by the \TESS Mission. These planets were found to be in or near the 2:1 resonance, and both of them exhibit transit timing variations (TTV). Precise characterization of the planets' masses and radii, orbital properties, and resonant behavior can test theories for the origins of planets orbiting close to their stars. Previous characterization of the system using the first six sectors of \TESS data suffered from a degeneracy between planet mass and orbital eccentricity. Radial velocity measurements using HARPS, FEROS, and PFS break that degeneracy, and an expanded TTV baseline from \TESS and an ongoing ground-based transit observing campaign increase the precision of the mass and eccentricity measurements. We determine that TOI-216c is a warm Jupiter, TOI-216b is an eccentric warm Neptune, and that they librate in the 2:1 resonance with a moderate libration amplitude of  $60^{+2}_{-2}$ degrees; small but significant free eccentricity of  $0.0222^{+0.0005}_{-0.0003}$ for TOI-216b; and small but significant mutual inclination of 1.2--3.9$^\circ$ (95\% confidence interval). The libration amplitude, free eccentricity, and mutual inclination imply a disturbance of TOI-216b before or after resonance capture, perhaps by an undetected third planet.
\end{abstract}

\section{Introduction}
Warm Jupiters -- defined here as giant planets with 10--100 day orbital periods -- in systems with other nearby planets are an important population for the investigation of giant planets close to their stars. High eccentricity tidal migration -- a strong contender for the origins of many close-in giant planets (see \citealt{daws18} for a review) is unlikely to have been at work in these systems (e.g., \citealt{must15}). Disk migration has been a persistently proposed explanation (e.g., \citealt{lee02}) -- particularly for systems in or near mean motion resonance -- but recent studies have argued that in situ formation may be more consistent with these planets' observed properties (e.g., \citealt{huan16,frel19,ande20}) and could possibly be consistent with resonant configurations (e.g.,  \citealt{dong16,macd18,chok20}). Precise characterization of the orbital properties and resonant behavior of individual systems can test origins scenarios, complementary to population studies.

One warm Jupiter system potentially amenable to such detailed characterization is TOI-216, which hosts a pair of warm, large exoplanets discovered by the \TESS Mission \citep{daws19,kipp19}. Their masses and orbits were characterized using transit timing variations (TTVs). However, previously the \TESS TTV dataset was not sufficiently precise to break the degeneracy between mass and eccentricity that arises when we can measure the near-resonant TTV signal but not the chopping TTV signal (e.g., \citealt{lith11,deck15}). Different priors on mass and eccentricity led to two qualitatively different solutions for the system \citep{daws19}: a Jupiter-mass planet accompanied by a Saturn-mass planet librating in orbital resonance, and a puffy sub-Saturn-mass planet and puffy Neptune-mass planet near but not in orbital resonance.  The solutions also differed in the planets' free eccentricity, the eccentricity not associated with the proximity to resonance. Since any origins scenario under consideration needs to be able to account for the planets' masses, free eccentricities, and resonant behavior, detailed characterization of these properties with a more extended dataset is warranted.

Since our previous study \citep{daws19}, seven more \TESS transits of planet b, three more \TESS transits of planet c, a ground-based transit of planet b, and five more ground-based transits of planet c have been observed, including recent ground-based transit observations that significantly extend the baseline for measuring transit timing variations beyond the observations conducted during the first year of the \TESS primary mission. Furthermore, we have been conducting a ground-based radial-velocity campaign using HARPS, FEROS, and PFS. Radial-velocity measurements serve to break the mass-eccentricity degeneracy.

Here we combine \TESS light curves, ground-based light curves, and ground-based radial velocities to precisely characterize  TOI-216b and c. In Section~\ref{sec:lc}, we describe our analysis of the \TESS data and the observation and analysis of ground-based light curves. We identify a weak stellar activity periodicity in the \TESS data that also shows up in the radial velocities. In Section~\ref{sec:rvs}, we present radial velocity measurements from HARPS, FEROS, and PFS and show that they immediately confirm the higher mass, lower eccentricity solution. We investigate additional weak periodicities in the radial-velocities and argue that they are caused by stellar activity. We jointly fit the transit timing variations and radial velocities in Section~\ref{sec:arch} and determine that the planet pair is librating in resonance, that the inner planet has a significant free eccentricity, and that planets have a small but significant mutual inclination. We present our conclusions -- including implications for the system's origins -- in Section~\ref{sec:discuss}.

\section{Light Curve Analysis}
\label{sec:lc}
This paper is based on data from TESS Sectors 1--13 (2018 July 25 -- 2019 July 17) and { Sectors 27--30 (2020 July 4 -- 2020 October 21)}, during which \thisstar\ was observed with CCD~1 on Camera~4, and from ground-based observatories. We use the publicly available 2-min cadence \TESS data, which is processed with the Science Processing Operations Center pipeline 
\citep{jenk16}. We download the publicly available data from the Mikulski Archive for Space Telescopes (MAST). The pipeline, a descendant of the \Kepler mission pipeline based at the NASA Ames Research Center \citep{jenk02,jenk10}, analyzes target pixel postage stamps that are obtained for pre-selected target stars. 

We use the resources of the \TESS Follow-up Observing Program (TFOP) Working Group (WG) Sub Group~1 (SG1)\footnote{\url{https://tess.mit.edu/followup/}} to collect seeing-limited time-series photometric follow-up of TOI-216.  All photometric time series are publicly available on the Exoplanet Follow-up Observing Program for TESS (ExoFOP-TESS) website\footnote{\url{https://exofop.ipac.caltech.edu/tess/}}. Light curves observed on or before February 24, 2019 are described in \citet{daws19}. Our new time-series follow-up observations are listed in Table~\ref{tab:ground}. We used the {\tt TESS Transit Finder}, which is a customized version of the {\tt Tapir} software package \citep{Jensen:2013}, to schedule our transit observations. Unless otherwise noted, the photometric data were extracted using the {\tt AstroImageJ} ({\tt AIJ}) software package \citep{Collins:2017}.  The facilities used to collect the new TOI-216 observations published here are Las Cumbres Observatory Global Telescope (LCOGT) network \citep{brown2013}, Hazelwood Observatory (Churchill, Vic, Australia), El Sauce Observatory (Coquimbo Province, Chile), and the Antarctic Search for Transiting ExoPlanets (ASTEP) observatory (Concordia Station, Antarctica). All LCOGT 1~m telescopes are equipped with the Sinistro camera, with a 4k$\times$4k pixel Fairchild back illuminated CCD and a $26\arcmin\times26\arcmin$ FOV. The LCOGT images were calibrated using the standard LCOGT BANZAI pipeline \citep{McCully:2018}. Hazelwood is a private observatory with an f/8 Planewave Instruments CDK12 0.32~m telescope and an SBIG STT3200 2.2k$\times$1.5k CCD, giving a $20\arcmin\times13\arcmin$ field of view. El Sauce is a private observatory which hosts a number of telescopes; the observations reported here were carried out with a Planewave Instruments CDK14 0.36~m telescope and an SBIG STT-1603-3 1536k$\times$1024k CCD, giving a $19\arcmin\times13\arcmin$ field of view. ASTEP is a 0.4~m telescope with an FLI Proline 16801E 4k$\times$4k CCD, giving a $63\arcmin\times63\arcmin$ field of view. The ASTEP photometric data were extracted using a custom IDL-based pipeline.

\begin{table*}
\footnotesize
\caption{Observation Log: TOI-216/TIC 55652896}         
\label{tab:ground}      
\centering 
\begin{tabular}{c l l c c c c c c c}  
\hline\hline       
\noalign{\smallskip}                  
\multirow{2}{*}{TOI-216} & Date & \multirow{2}{*}{Telescope}\tablenotemark{$\dag$} & \multirow{2}{*}{Filter} & ExpT & Exp & Dur. & Transit & Ap. Radius  & FWHM\\
& (UTC) &  &  & (sec) & (N) & (min) &Coverage & (arcsec) & (arcsec)   \\
\noalign{\smallskip} 
\hline  
\noalign{\smallskip}                  
\multirow{1}{*}{\inn}
& 2019-10-30 & LCOGT-CTIO-1.0  & I          & 40  & 220 & 241  & Full           & 5.8  & 2.4 \\
& 2020-05-23 & ASTEP-0.4     & $\sim$Rc   & 120 &376& 924&Full&10.2&5.3\\ 
\hline
\noalign{\smallskip} 
\multirow{2}{*}{\out}
& 2018-12-16 & LCOGT-SAAO-1.0  & i$^\prime$ & 90  & 331 & 450  & Full           & 5.8  & 2.1 \\
& 2019-01-20 & Hazelwood-0.3 & g$^\prime$ & 240 & 101 & 449  & Egress+70$\%$  & 5.5  & 3.2 \\
& 2019-02-24 & El Sauce-0.36 & Rc         & 30  & 514 & 303  & Egress+90$\%$  & 5.9  & 3.4 \\
& 2019-06-07 & ASTEP-0.4     & $\sim$Rc   & 120 & 549 & 1440 & Full           & 12.0 & 5.0 \\
& 2019-11-27 & Hazelwood-0.3 & Rc         & 60  & 114 & 255  & Egress+60$\%$  & 5.5  & 3.1 \\
& 2019-12-31 & LCOGT-SAAO-1.0  & Ic         & 40  & 316 & 345  & Ingress+90$\%$ & 4.7  & 2.0 \\
& 2020-02-24 & Hazelwood-0.3 & Rc         & 120 & 161 & 380  & Egress+60$\%$  & 5.5  & 3.1 \\
& 2020-03-09 & LCOGT-CTIO-1.0  & I          & 40  & 146 & 176  & Egress+30$\%$  & 8.2  & 2.6 \\
& 2020-06-21 & ASTEP-0.4     & $\sim$Rc   & 120 & 272 & 656  & Full           & 11.0 & 5.0 \\
\noalign{\smallskip}
\hline
\noalign{\smallskip}
\end{tabular}
\tablenotetext{$\dag$}{Telescopes: \\
             LCOGT-CTIO-1.0: Las Cumbres Observatory - Cerro Tololo Inter-American Observatory (1.0 m) \\
             LCOGT-SAAO-1.0: Las Cumbres Observatory - South African Astronomical Observatory (1.0 m) \\
             LCOGT-SSO-1.0: Las Cumbres Observatory - Siding Spring Observatory (1.0 m) \\
             Hazelwood-0.3: Stockdale Private Observatory - Victoria, Australia (0.32 m) \\
             El Sauce-0.36: El Sauce Observatory - Coquimbo Province, Chile (0.36 m) \\
             ASTEP-0.4: Antarctic Search for Transiting ExoPlanets - Concordia Station, Antarctica (0.4 m) \\
             }
\end{table*}

\begin{figure*}
\begin{center}
\includegraphics{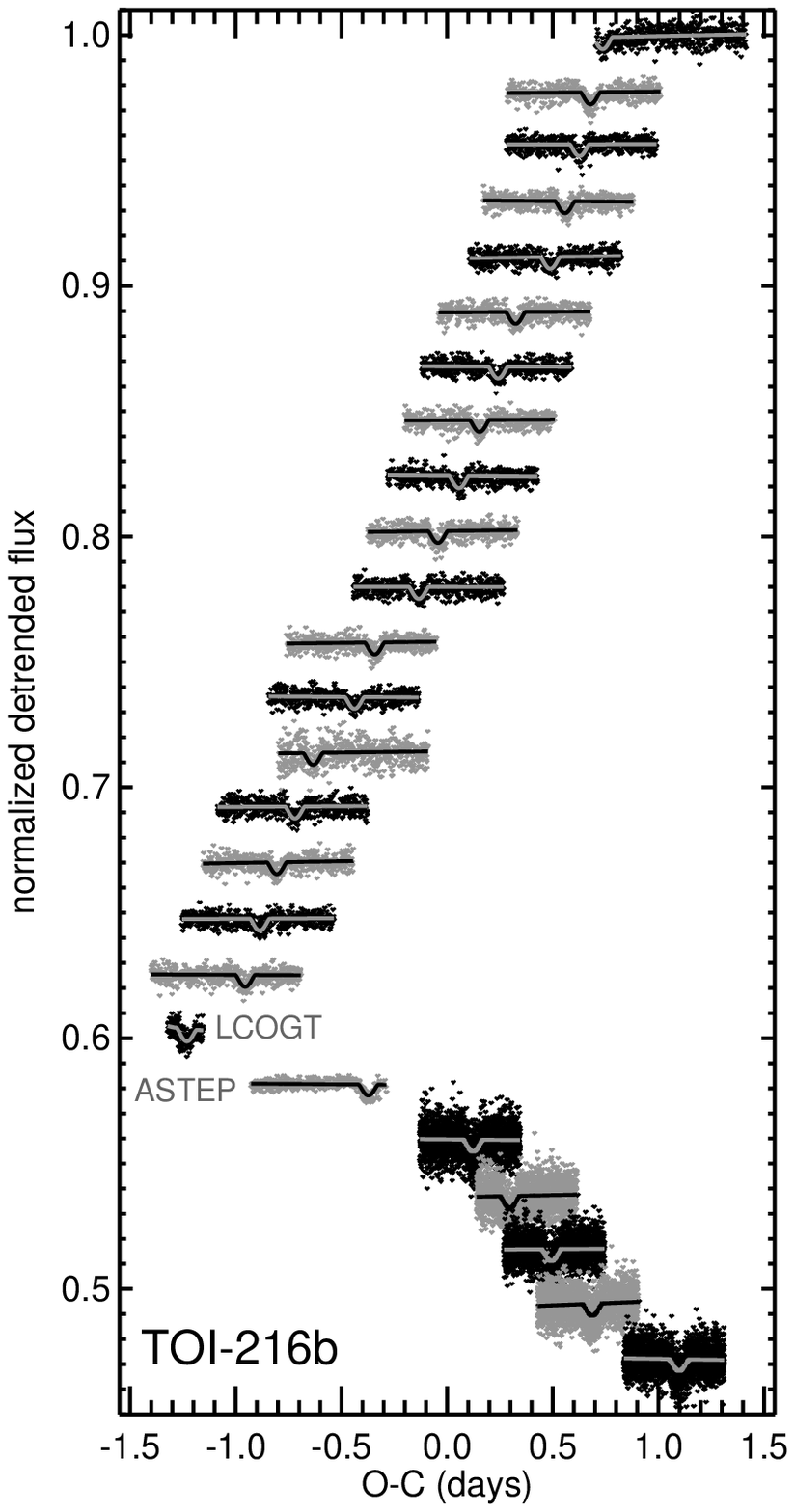}
\includegraphics{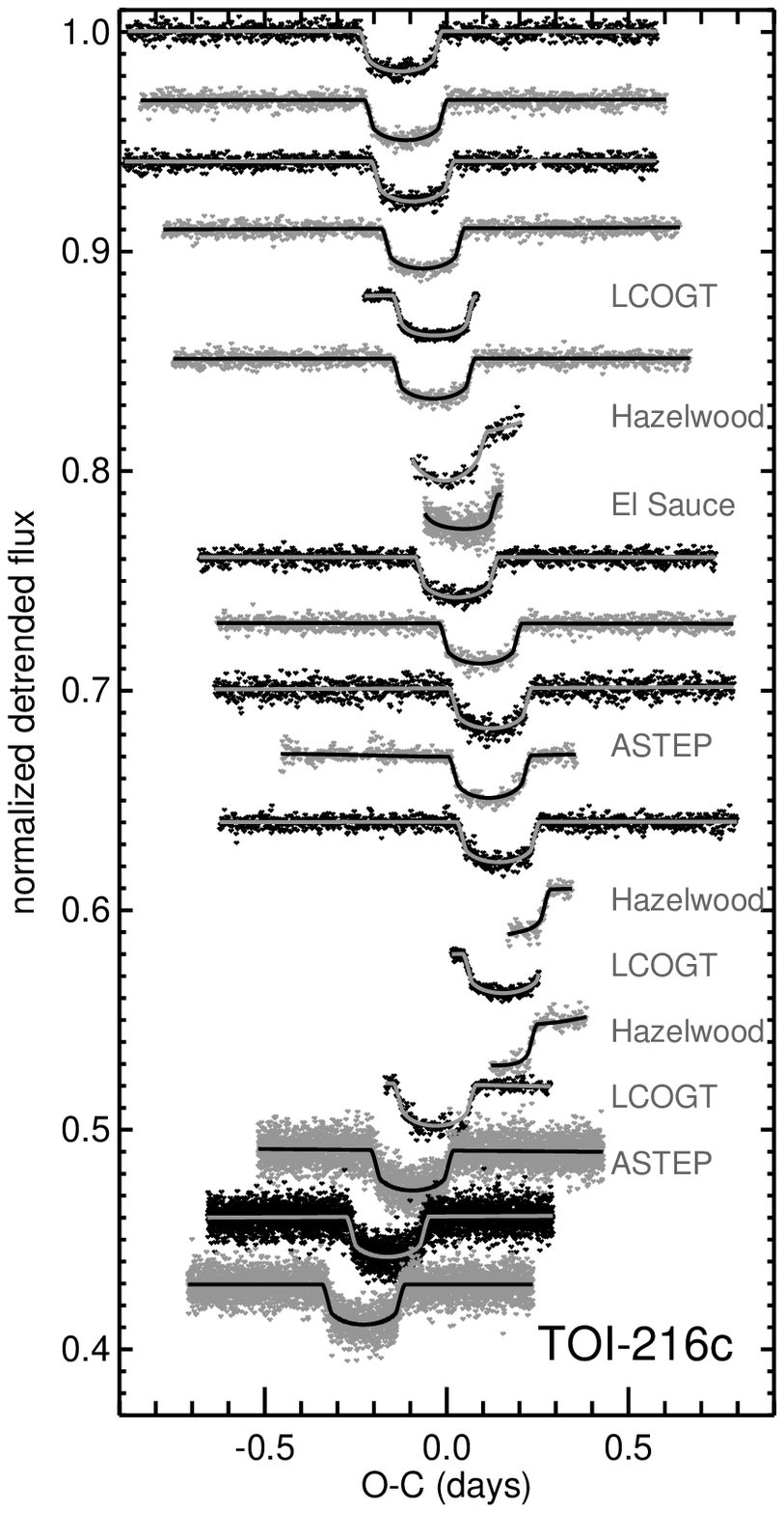}
\caption{ 
 \label{fig:transits} 
 Detrended light curves, spaced with arbitrary vertical  offsets, and with a model light curve overplotted. The light curves are phased based on a constant orbital period linear ephemeris to show the TTVs. TESS data are publicly available from MAST and ground-based data from the ExoFOP-TESS website.}
\end{center}
\end{figure*}

 We fit the transit light curves (Fig. \ref{fig:transits}) using the TAP software package \citep{gaza12}, which implements Markov Chain Monte Carlo using the \citet{mand02} transit model and the \citet{cart09} wavelet likelihood function, with the modifications described in \citet{daws14}. The results are summarized in Table~\ref{tab:216lc}. For \TESS light curves, we use the presearch data conditioned (PDC) flux, which is corrected for systematic (e.g., instrumental) trends using cotrending basis vectors \citep{smit12,stum14}. For all light curves, we use \citet{cart09} wavelet likelihood function (which, for the red noise component, assumes noise with an amplitude that scales as  frequency$^{-1}$) with free parameters for the amplitude of the red $\sigma_r$ and white noise $\sigma_r$ and a linear trend fit simultaneously to each transit light curve segment with other transit parameters. For the ground-based observations, we fit a linear trend to airmass instead of time.  We assign each instrument and filter (\TESS, Hazelwood g' and Rc, LCOGT i' and I, El Sauce Rc, and ASTEP Rc)  its own set of limb darkening parameters because of the different wavebands. We use one set of noise parameters for all the \TESS light curves and an additional set for each ground-based light curve. We adopt uniform priors on the planet-to-star radius ratio ($R_{p}/R_{\star}$), the impact parameter $b$ (which can be either negative or positive; we report $|b|$), the mid transit time, the limb darkening coefficients $q_1$ and $q_2$ \citep{kipp13}, and the slope and intercept of each transit segment's linear trend. For the grazing transits of the inner planet, we impose a uniform prior on $R_p/R_\star$ from 0 to 0.17, with the upper limit corresponding to a planet radius of 0.13 solar radii (see \citealt{daws19} for details and justification). Despite a well-constrained transit depth (Table \ref{tab:216lc}), the inner planet's radius ratio is highly uncertain due to degeneracy between the radius ratio and impact parameter (see Fig. 3 of \citealt{daws19}). We also impose a uniform prior on the log of the light curve stellar density $\rhocirc$. We use $\rhocirc$, the stellar density derived from the light curve assuming a circular orbit, to compute the \citet{mand02} model normalized separation of centers $z = d/R_\star$, assuming $M_p << M_\star$ and a circular orbit:
 \begin{eqnarray}
 \label{eqn:rhocirc}
 z = \left(\rhocirc/\rho_\odot\right)^{1/3} \left(P/P_\oplus\right)^{2/3} \left(a_\oplus/R_\odot\right)
 \end{eqnarray}
where $P$ is the orbital period, the subscript $\oplus$ denotes the Earth, and the subscript $\odot$ denotes the Sun. We will later combine the posteriors $\rhocirc$ from the light curve and $\rho_\star$ (from \citealt{daws19}'s isochrone fit) to constrain the orbital eccentricity (Section \ref{subsec:joint}). { We perform an additional set of fits where we allow for a dilution factor for the TESS light curves and find the results to be indistinguishable. We measure a dilution factor of $1.000^{+0.012}_{-0.11}$.}

 We also perform two additional fits to look for transit duration variations, following \citet{daws20}. In the first fit, we allow the impact parameter to vary with the prior recommended by \citet{daws20}. { We find tentative evidence for the variation in the transit impact parameter for the inner planet, with an impact parameter change scale of 0.007$^{+0.004}_{-0.003}$.}  In the second fit, we allow $\rhocirc$ (Eqn. \ref{eqn:rhocirc}) and $b$ to vary for each transit with a prior that corresponds to a uniform prior in transit duration \citep{daws20}. { Again the inner planet exhibits tentative evidence for variation in its transit durations.} We fit a line to the transit durations as a function of mid-transit time and { determine a three-sigma confidence interval on the slope of -0.4 -- 4 seconds/day for the inner planet and -1 -- 0.4 seconds/day for the outer planet}.
 
 For comparison and to ensure that the results are not sensitive to the correlated noise treatment, we also fit the light curves using the \texttt{exoplanet} package \citep{exoplanet:exoplanet}, which uses Gaussian process regression. Each set of \TESS light curves along with eight ground-based light curves is modeled with a light curve transit model built from \texttt{starry} \citep{exoplanet:luger18} plus a Matern-3/2 Gaussian process kernel with a white noise term. Seven sets of limb-darkening parameters \citep{exoplanet:kipping13} are used for observations conducted in seven different filters. We use a log-uniform prior on stellar densities ($\rho_{\rm circ}$), log-normal prior on transit depths, uniform prior on the impact parameters, and uniform prior on mid-transit times. We infer posteriors for each parameter using this approach that are consistent within 1 sigma to our nominal fit above.
 \clearpage
 \startlongtable
\begin{deluxetable}{rrl}
\tabletypesize{\footnotesize}
\tablecaption{ Planet Parameters for \thisstarinn and \thisstarout Derived from the Light Curves \label{tab:216lc}}
\tablewidth{0pt}
\tablehead{
\colhead{Parameter}    & \colhead{Value\tablenotemark{a}}}
\startdata
\hline
\thisstarinn\\
Planet-to-star radius ratio, $R_{p}/R_{\star}$   				&0.10 &$^{+0.03}_{-0.02}$ \\
Transit depth [ppm] & 4750&$^{+140}_{-150}$\\
Planet radius, $R_p$ [$R_\oplus$] & 8 & $^{+3}_{-2}$\\
Light curve stellar density\tablenotemark{b}, $\rhocirc$  [$\rho_\odot$]  					&1.1 &$^{+0.3}_{-0.2}$ \\
Impact parameter, $|b|$ 	&	0.98 &$^{+0.05}_{-0.04}$\\
Mid-transit times (days\tablenotemark{c,d}) & 1325.328 &$^{+0.004}_{-0.004}$\\
&1342.430 &$^{+0.003}_{-0.003}$\\ 
&1359.540 &$^{+0.003}_{-0.003}$\\
&1376.631 &$^{+0.003}_{-0.003}$\\
&1393.723 &$^{+0.003}_{-0.003}$\\
&1427.879 &$^{+0.003}_{-0.003}$\\
&1444.958 &$^{+0.003}_{-0.003}$\\
&1462.031 &$^{+0.003}_{-0.003}$\\
&1479.094 &$^{+0.003}_{-0.003}$\\
&1496.155 &$^{+0.003}_{-0.003}$\\
&1513.225 &$^{+0.004}_{-0.004}$\\
&1547.338 &$^{+0.003}_{-0.003}$\\
&1564.403 &$^{+0.004}_{-0.004}$\\
&1598.529 &$^{+0.003}_{-0.003}$\\
&1615.604 &$^{+0.003}_{-0.003}$\\
&1632.679 &$^{+0.003}_{-0.003}$\\
&1649.759 &$^{+0.004}_{-0.004}$\\
&1666.851 &$^{+0.003}_{-0.003}$\\
LCOGT&1786.698 &$^{+0.002}_{-0.002}$\\
{ ASTEP}&1993.486&$^{+0.002}_{-0.002}$\\
&{ 2045.465}&$^{+0.003}_{-0.003}$\\ 
&{ 2062.800}&$^{+0.003}_{-0.003}$\\ 
&{ 2080.157}&$^{+0.003}_{-0.003}$\\ 
&{ 2097.511}&$^{+0.003}_{-0.003}$\\ 
&{ 2132.243}&$^{+0.003}_{-0.003}$\\ 
Best fit linear ephemeris:\\
Period (days) &{ 17.16073 }\\
Epoch (days) & { 1324.5911}\\
\hline
\thisstarout \\
Planet-to-star radius ratio, $R_{p}/R_{\star}$   				&0.1230 &$^{+0.0008}_{-0.0006}$ \\
Transit depth [ppm] & 18310&$^{+120}_{-120}$\\
Planet radius, $R_p$ [$R_\oplus$] & 10.1 & $^{+0.2}_{-0.2}$\\
Light curves stellar density, $\rhocirc$  [$\rho_\odot$]  					&1.73 &$^{+0.04}_{-0.08}$ \\
Impact parameter, $|b|$ 					&0.14 &$^{+0.08}_{-0.09}$ \\
Mid-transit times (days\tablenotemark{c,d}) & 1331.2850 &$^{+0.0008}_{-0.0008}$\\
&1365.8244 &$^{+0.0008}_{-0.0008}$\\ 
&1400.3686 &$^{+0.0008}_{-0.0008}$\\ 
&1434.9227&$^{+0.0008}_{-0.0008}$\\ 
&1469.4773&$^{+0.0008}_{-0.0008}$\\   
LCOGT&1469.4781&$^{+0.0004}_{-0.0004}$\\   
Hazelwood&1504.036 &$^{+0.002}_{-0.002}$\\  
El Sauce&1538.5938 &$^{+0.0015}_{-0.0015}$\\  
&1538.5921&$^{+0.0008}_{-0.0008}$\\  
&1607.7082&$^{+0.0008}_{-0.0008}$\\  
&1642.2612&$^{+0.0009}_{-0.0009}$\\  
{ ASTEP}&1642.2595&$^{+0.0011}_{-0.0011}$\\
&1676.8085&$^{+0.0008}_{-0.0008}$\\  
Hazelwood&1814.939&$^{+0.003}_{-0.003}$\\  
LCOGT&1849.4526&$^{+0.0005}_{-0.0005}$\\  
Hazelwood&1883.955&$^{+0.002}_{-0.002}$\\
LCOGT&1918.4506&$^{+0.0007}_{-0.0007}$\\  
ASTEP&2021.8887&$^{+0.0010}_{-0.0010}$\\
&{ 2056.3520}&$^{+0.0008}_{-0.0008}$\\ 
&{ 2090.8107}&$^{+0.0009}_{-0.0009}$\\ 
&{ 2125.2698}&$^{+0.0008}_{-0.0008}$\\ 
Best fit linear ephemeris:\\
Period (days) &{ 34.525528}\\
Epoch (days) & { 1331.4110}\\
\enddata
\tablenotetext{a}{As a summary statistic we report the median and 68.3\% confidence interval  of the posterior distribution.}
\tablenotetext{b}{Eqn. \ref{eqn:rhocirc}}
\tablenotetext{c}{BJD - 2457000.0 days}
\tablenotetext{d}{Mid-transit times not otherwise noted are from \TESS light curves.}
\end{deluxetable}
\clearpage

\begin{figure}
\begin{center}
\includegraphics{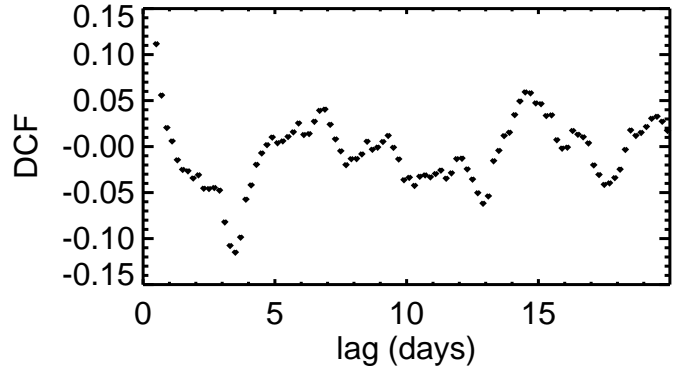}
\caption{
 \label{fig:dcf}
Discrete autocorrelation function (Eqn. 2 of \citealt{ede88}) of {\tt eleanor} long cadence TOI-216 light curve.}
\end{center}
\end{figure}

We examine the light curve for evidence of stellar rotation. We do not see any significant periodicities in the SPOC 2 minute cadence data. To investigate further, we create a systematics-corrected long cadence light curve using {\tt eleanor} \citep{fein19} with a 15 x 15 target pixel file, background size of 100, and custom square aperture of 3x3 pixels centered on TOI-216. We follow {\tt eleanor}'s recommendation for background subtraction: the 1D postcard background for sectors 1, 2, 4, 6, 7, 8, 11 and 13; the 1D target pixel file background for sectors 3 and 5; and the 2D target pixel file background for sectors 9 and 12. Then we mask out the transits and compute a discrete autocorrelation function (DCF; Eqn. 2 of \citealt{ede88}), plotted in Fig. \ref{fig:dcf}. 

The peak at 6.5 day and valley at approximately half that value are consistent with a 6.5 day periodicity. This periodicity could represent the rotation period or a shorter harmonic. A rotation period of $\sim 10-40$ days would be most typical for a 0.78 $M_\odot$ star on the main sequence (e.g., \citealt{mcqu14}), so the periodicity could plausibly be an integer fraction of the rotation period (e.g., one half, one third). The stellar radius of 0.747$^{+0.015}_{-0.014} R_\odot$ \citep{daws19} and  $v \sin i = 0.84 \pm 0.70$ km/s from the FEROS spectra correspond to a rotation period of $45^{+225}_{-20}$ days assuming an edge-on orientation. A rotation period of 26 days would be compatible with the $v \sin i$ measurement, or a shorter rotation period would indicate a spin-orbit misalignment. A periodicity of 13 days or 26 days is challenging to detect in the \TESS light curve (e.g., \citealt{cant20}). 13 days is close to \TESS's 13.7 day orbital period and thus susceptible to corrections for systematics. Both are a significant fraction of the sector (27 days) and subject to imprecision in stitching together different segments. Given the many bumps and wiggles in the DCF, we do not consider this lightcurve detection of a  stellar rotation harmonic definitive. 

\begin{deluxetable}{rrlrlrlrl}
\tabletypesize{\footnotesize}
\tablecaption{ Light Curve Parameters\tablenotemark{a} for the \thisstar system\label{tab:216star}}
\tablewidth{0pt}
\tablehead{
\colhead{}    & \colhead{$q_1$}&&\colhead{$q_2$}&&\colhead{$\sigma_r$ }&\colhead{[ppm]}&\colhead{$\sigma_w$}&\colhead{[ppm]}}
\startdata
TESS & $0.29$& $^{+0.08}_{-0.05}$ & $0.50$&$^{+0.11}_{-0.10}$ \\
&2 min & 3900 & $^{+600}_{-600}$ & 2475&$^{+14}_{-14}$ \\
&20 s  & 15000 & $^{+2000}_{-2000}$ &  5364&$^{+26}_{-27}$ \\
LCOGT i' & $0.55$& $^{+0.10}_{-0.10}$ & $0.20$&$^{+0.07}_{-0.07}$ & 1600 & $^{+1970}_{-1100}$ & 1060&$^{+40}_{-40}$ \\
Hazelwood g'  & $0.51$& $^{+0.25}_{-0.15}$ &$0.6$& $^{+0.2}_{-0.2}$ & 4000 & $^{+3000}_{-3000}$ & 2450&$^{+190}_{-180}$ \\
El Sauce  & $0.5$& $^{+0.2}_{-0.2}$ & $0.3$&$^{+0.2}_{-0.2}$ & $10^4$ & $^{+4000}_{-4000}$ & 3130&$^{+80}_{-80}$ \\
Hazelwood Rc:& $0.6$& $^{+0.3}_{-0.4}$ &$0.11$& $^{+0.18}_{-0.08}$ \\
Nov. 2019&&&&& 9000 & $^{+3000}_{-2000}$ & 2100&$^{+200}_{-200}$ \\
Feb. 2020&&&&&  $10^4$ & $^{+3000}_{-3000}$ & 1920&$^{+160}_{-160}$ \\
LCOGT I: & $0.29$& $^{+0.07}_{-0.09}$ & $0.49$& $^{+0.19}_{-0.10}$ \\
Oct. 2019&&&&& 1400 & $^{+1500}_{-1100}$ & 2360&$^{+110}_{-100}$ \\
Dec. 2019&&&&& 1200 & $^{+1100}_{-800}$ & 1130&$^{+40}_{-40}$ \\
Mar. 2020&&&&& 1600 & $^{+1400}_{-1100}$ & 990&$^{+50}_{-50}$ \\
ASTEP& $0.52$& $^{+0.18}_{-0.13}$ & $0.4$& $^{+0.2}_{-0.2}$ \\
June 2019&&&&& 13000 & $^{+2000}_{-2000}$ & 2100&$^{+100}_{-100}$ \\
May 2020&&&&&3800 & $^{+1300}_{-1100}$ & 1270&$^{+50}_{-50}$ \\
June 2020&&&&&11900 & $^{+1500}_{-1500}$ & 1360&$^{+70}_{-70}$ \\
\enddata
\tablenotetext{a}{As a summary statistic we report the mode and 68.3\% confidence interval  of the posterior distribution.}
\end{deluxetable}

\section{Radial velocity analysis}
\label{sec:rvs}

TOI-216 was monitored with three different high resolution echelle spectrographs over a time span of 16 months with the goal of obtaining precision radial velocities to further constrain the masses and orbital parameters of the giant planets present in the TOI-216 system. These observations were performed in the context of the Warm gIaNts with tEss (WINE) collaboration, which focuses on the systematic characterization of TESS transiting giant planets with orbital periods longer than $\approx$10 days \citep[e.g.,][]{brahm:2019,jordan:2020}. All radial velocity measurements of TOI-216 are presented in Table~\ref{tab:rvs}.

We obtained 27 spectra with the Fibre-fed, Extended Range, Échelle Spectrograph \citep[FEROS,][]{kaufer:99} between November of 2018 and March of 2019. FEROS is mounted on the MPG~2.2~m telescope at the ESO La Silla Observatory, and has a resolving power of R\,$\approx$\,48\,000. Observations were performed with the simultaneous calibration mode for tracing radial velocity variations produced by environmental changes in the instrument enclosure. The adopted exposure time of 1200 seconds yielded spectra with signal-to-noise ratios in the range from 40 to 75. FEROS spectra were processed from raw data with the \texttt{ceres} pipeline \citep{ceres}, which delivers precision radial velocities and line bisector span measurements via cross-correlation with a binary mask resembling the spectral properties of a G2-type star. The radial velocity uncertainties for the FEROS observations of TOI-216 range between 7 and 15 m s$^{-1}$. { We remove two outliers from the FEROS radial-velocity time series at 1503.75 and 1521.57 days. All subsequent results do not include these outliers. We have checked that no results except the inferred jitter for the FEROS dataset are sensitive to whether or not the outliers are included.}

We observed TOI-216 on 15 different epochs between December of 2018 and October of 2019 with the High Accuracy Radial velocity Planet Searcher \citep[HARPS][]{harps} mounted on the ESO~3.6~m telescope at the ESO La Silla Observatory, in Chile. We adopted an exposure time of 1800 seconds for these observations using the high radial velocity accuracy mode (HAM, R\,$\approx$\,115\,000), which produced spectra with signal-to-noise ratios of $\approx$40 per resolution element. As in the case of FEROS, HARPS data for TOI-216 was processed with the \texttt{ceres} pipeline, delivering radial velocity measurements with typical errors of $\approx$5 m s$^{-1}$.

TOI-216 was also monitored with the Planet Finder Spectrograph \citep[PFS;][]{craneetal2006,craneetal2008,Craneetal2010} mounted on the 6.5\,m Magellan II Clay Telescope at Las Campanas Observatory (LCO), in Chile. These spectra were obtained on 18 different nights, between December of 2018 and March of 2020, using the 0.3$\arcsec\times$2.5$\arcsec$ slit, which delivers a resolving power of $R$\,=\,130\,000. Due to its moderate faintness, TOI-216 was observed with the 3$\times$3 binning mode to minimize read-out noise, and an exposure time of 1200 seconds was adopted to reach a radial velocity precision of $\approx$2 m s$^{-1}$. An iodine cell was used in these observations as a reference for the wavelength calibration. The PFS data were processed with a custom IDL pipeline \citep{Butler1996}. Three consecutive 1200 second iodine-free exposures of TOI-216 were obtained to construct a stellar spectral template for disentangling the iodine spectra from the stellar one for computing the radial velocities.

\begin{figure*}
\begin{center}
\includegraphics{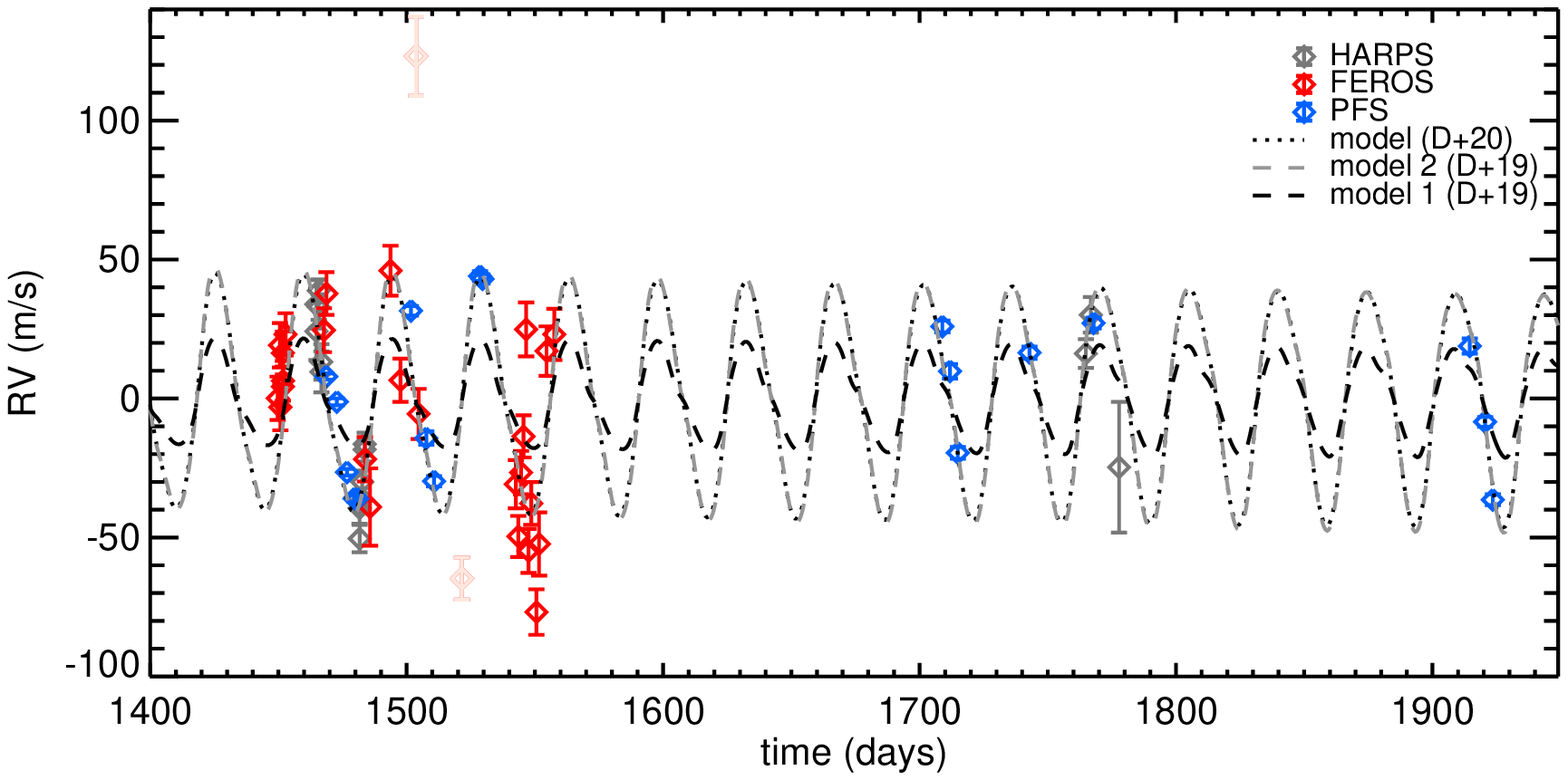}
\includegraphics{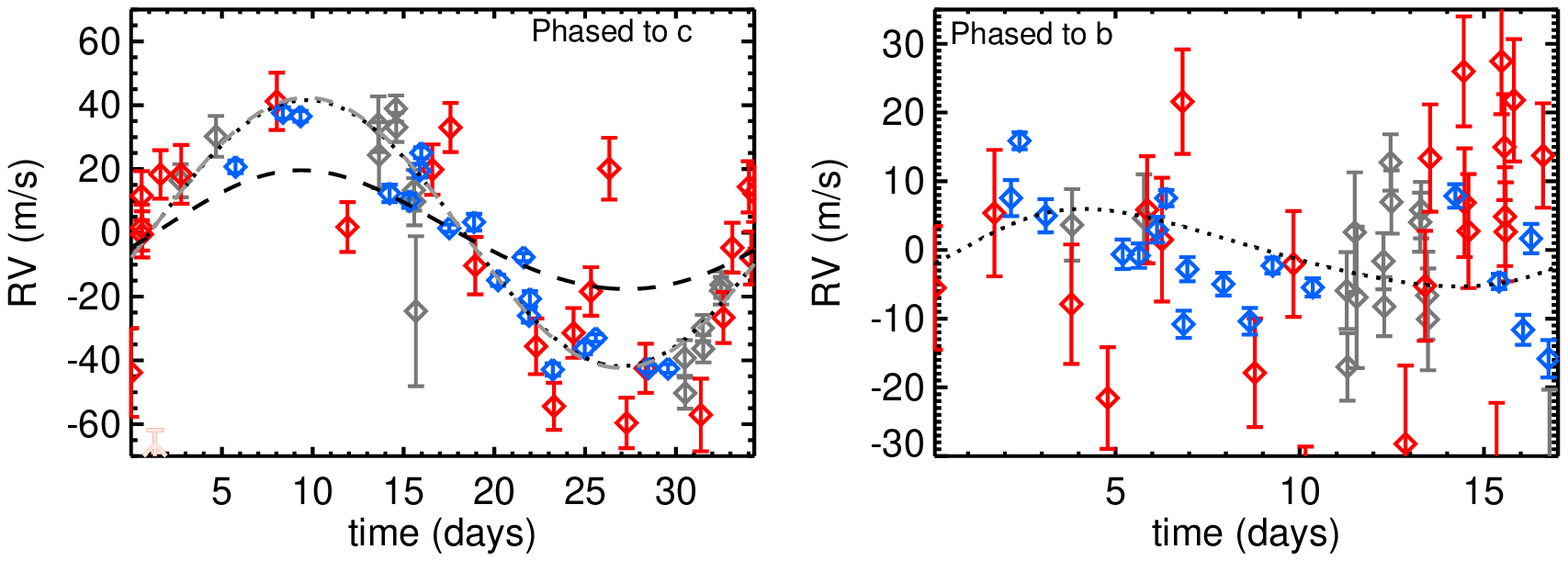}
\caption{
 \label{fig:rv}
Top: Radial velocity measurements of TOI-216 from { HARPS (gray), FEROS (red; outliers in lighter red), and PFS (blue)}. Best-fit models from this work (Table \ref{tab:216}) and from \citet{daws19}'s earlier analysis of transit times only are overplotted. Row 2, left: RVs phased to planet c's orbital period with same models as row 1 for planet c only (i.e., the RV variation only due to planet c) overplotted. Row 2, right: Residuals of Table \ref{tab:216} model for planet c only, with b component of the model overplotted, phased to planet b's orbital period.} 
\end{center}
\end{figure*}

\begin{figure*}
\begin{center}
\includegraphics{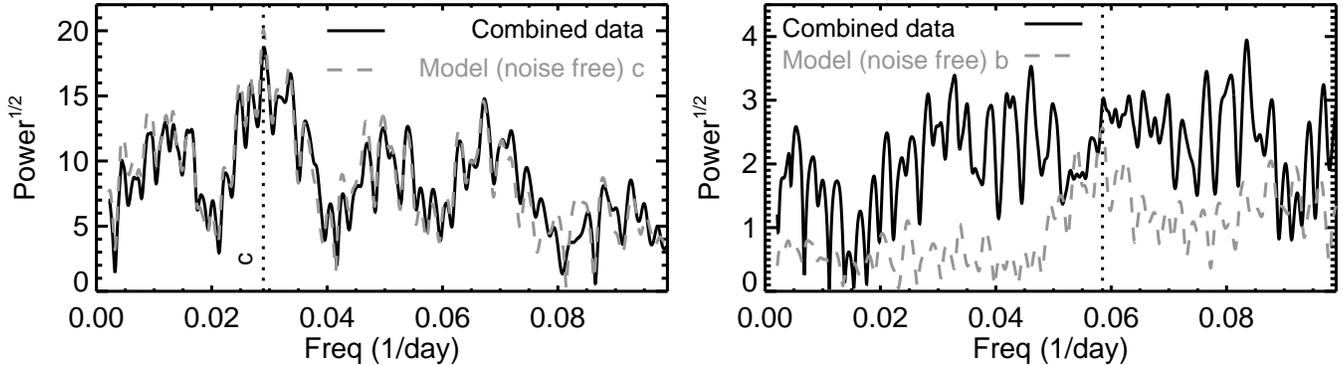}
\caption{
 \label{fig:rvper}
 Left: Periodograms of the combined RV dataset (black solid), with periodogram of noise-free planet c only model ( i.e., a Keplerian signal computed using the parameters of planet c from Table \ref{tab:216} sampled at the observed times in Table \ref{tab:rvs}; gray dashed) overplotted. Right: Same for residuals to planet c only model, with noise-free planet b only model overplotted  (i.e., a Keplerian signal computed using the parameters of planet b from Table \ref{tab:216} sampled at the observed times in Table \ref{tab:rvs}).}
\end{center}
\end{figure*}

\begin{figure}
\begin{center}
\includegraphics{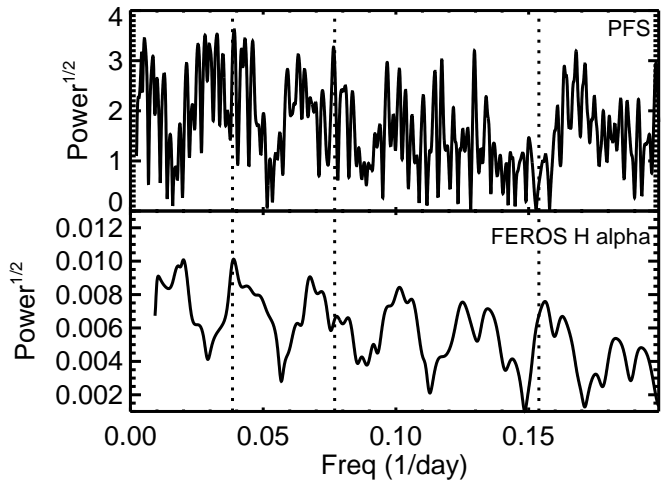}
\caption{{
 \label{fig:resid}
Periodograms of residuals of the best two planet fit (Table \ref{tab:216}) for PSF  and H alpha for FEROS.} The 6.5, 13, and 26 day periodicities  (associated with the periodicity in the light curve, Fig. \ref{fig:dcf}) are overplotted as dotted lines.}
\end{center}
\end{figure}

It is immediately evident that the radial velocities show good agreement with \citet{daws19}'s higher mass solution, which was fit to the earlier, transit time only dataset (Fig. \ref{fig:rv}), and with \citet{kipp19}'s solution based on transit times from the first six sectors. The bottom panel of Fig.~\ref{fig:rv} shows the RVs phased to the outer planet's orbital period. { We compute a generalised Lomb-Scargle periodogram \citep{cumm99,zech09} in Fig. \ref{fig:rvper} and \ref{fig:resid}. The x-axis is frequency $f$ in cycles per day. The y-axis is the square root of the power, where we define power as}
\begin{equation}
    {\rm Power}_f = \frac{\Sigma_i \frac{\left(v_{\rm i, f}-v_{i,0}\right)^2}{\sigma_i^2} - \Sigma_i \frac{v_{i,0}^2}{\sigma_i^2}}{2 \Sigma_i \frac{1}{\sigma_i^2}},
\end{equation}
{ where $\sigma_i$ is the reported uncertainty (Table \ref{tab:rvs}) and $v_{i,0}$ is the radial-velocity (Table \ref{tab:rvs}) with the error weighted mean $\frac{\Sigma_i v_i/\sigma_i^2}{\frac{1}{\sigma_i^2}}$ for each of the three datasets (PFS, HARPS, and FEROS) subtracted. The sinusoidal function $v_{\rm i, f}$ is }
\begin{equation}
    v_{\rm i, f} = A \cos[2\pi f(t_i-t_0)] + BA \sin[2\pi f(t_i-t_0)] + C_k
\end{equation}
{ where $A$ and $B$ are computed following \cite{zech09}, $k$ is each of the three datasets, $t_0$ is the time of the first observed radial velocity, and}
\begin{equation}
   C_k = -\frac{ \Sigma_i \frac{(A \cos[2\pi f(t_i-t_0)] + BA \sin[2\pi f(t_i-t_0)])}{\sigma_i^2}}{ \Sigma_i \frac{1}{\sigma_i^2}}.
\end{equation}

{ We see a peak in the periodogram at planet c's orbital period (Fig.~ \ref{fig:rvper}, left panel). The noise-free c model (overplotted, dashed; a Keplerian signal computed using the parameters of planet c from Table \ref{tab:216} sampled at the observed times in Table \ref{tab:rvs}) shows that many of the other peaks seen in the three periodograms are aliases of planet c's orbital period ( i.e., they caused by the observational time sampling of the planet's signal). Planet b's signal is below the noise level (Fig.~\ref{fig:rv}, bottom right panel; Figure~\ref{fig:rvper}, right panels).} In the PFS dataset, including planet b improves the chi-squared from 403 to 328 (for 18 datapoints and five additional parameters); for the HARPS and FEROS datasets, there is no improvement in chi-squared. Given that planet b is barely detected, we cannot rule out other planets in the system with smaller orbital radial velocity amplitudes.

Systems containing an outer planet in or near a 2:1 mean motion resonance with a less massive inner planet can be mistaken for a single eccentric planet, because the first eccentricity harmonic appears at half the  planet's orbital period (e.g., \citealt{angl10,kurs15}). In the case of the system TOI-216, the lack of detection of TOI-216b is not due to this phenomenon because solution we subtract off for planet c has near 0 eccentricity. However, without prior knowledge that the system contains a resonant pair, if we only had the radial-velocity datasets and the datasets were less noisy (or had more data points), we might be sensitive to planet b's signal but mistakenly interpret it as planet c's eccentricity.

\begin{figure}
\begin{center}
\includegraphics{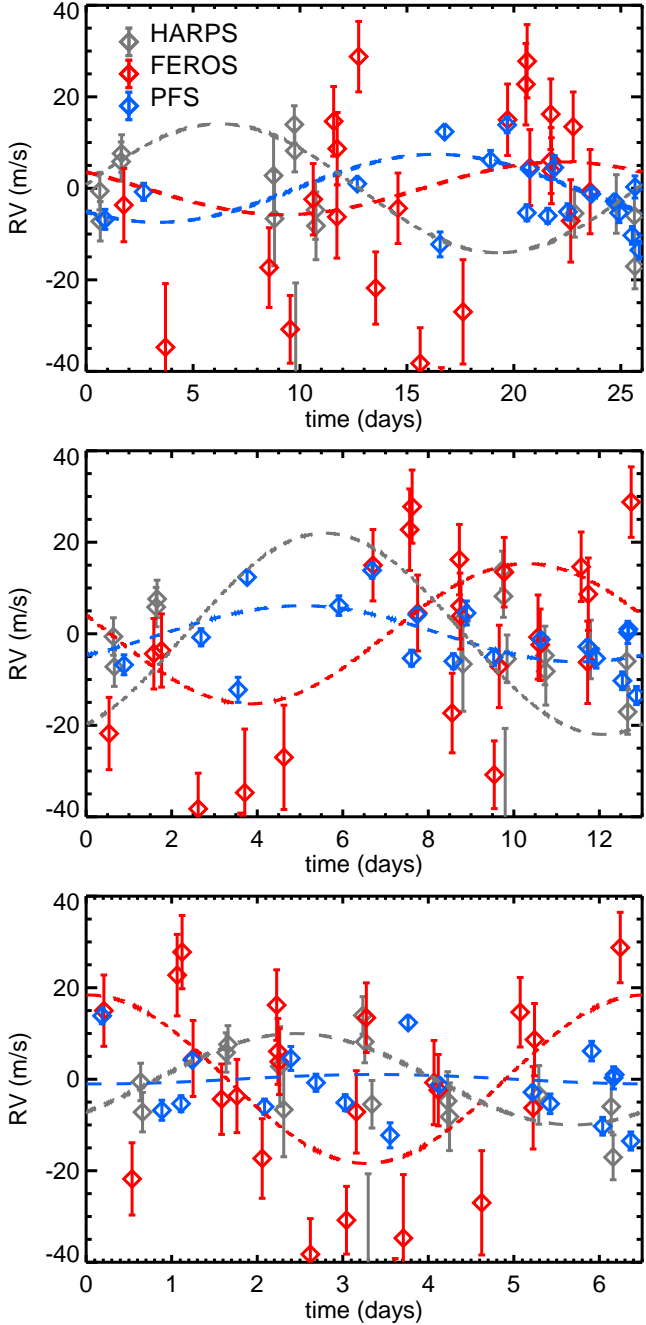}
\caption{
 \label{fig:residphase}
Residuals to the best two planet fit (Table \ref{tab:216}) phased to periodicities of 26 days (top), 13 days (middle), and 6.5 days (bottom).  Best-fit sinusoids for these periodicities are overplotted as dashed lines. { Comparing the three datasets, the best-fit sinusoids are out of phase with each other, indicating that these signals are unlikely to arise from additional planets.} }
\end{center}
\end{figure}

The residuals of the two-planet fit (and one-planet fit) show evidence of a signal that we attribute to stellar activity. { We examined the} periodograms of the residuals of each of three datasets; the bisectors for the HARPS and FEROS datasets; and H alpha for the FEROS dataset, computed with the {\tt ceres} pipeline, following \citet{bois09}. { Periodograms of the PFS residuals and FEROS H alpha are shown in Figure \ref{fig:resid}. There are no strong peaks in any of the residual or activity indicator periodograms.}  Some residuals datasets exhibit (weak) peaks near the 6.5 day periodicity (HARPS residuals and bisectors; FEROS residuals) identified in the light curve (Section \ref{sec:lc}), or a multiple of 6.5 days (PFS; FEROS bisectors and H alpha). In Fig. \ref{fig:residphase}, we plot the residuals of the two-planet fit with best-fit sinusoids for 6.5, 13, and 26 day periodicities for each dataset. Comparing the three datasets, best-fit sinusoids are out of phase with each other. The similarity of the periodicities to those seen in the light curve, the appearance of the 6.5 day periodicity in the HARPS bisectors, the 6.5 and 26 day periodicities in the FEROS H alpha, and the 13 day and 6.5 day periodicities in the FEROS bisectors, and the difference in phase among datasets suggest the { weak periodicities do not arise from additional planets. They are possibly caused by stellar variability. }

\section{Joint fit and orbital architecture}
\label{sec:arch}

Here we jointly fit the transit light curves (Section \ref{sec:lc}) and radial-velocity measurements (Section \ref{sec:rvs}) to precisely measure the orbital parameters and masses of both planets. In Section \ref{subsec:ttv}, we describe our analysis of the transit timing measurements. In Section \ref{subsec:joint}, we jointly fit the transit and radial-velocity measurements.

\subsection{Transit timing variation analysis}
\label{subsec:ttv}

\begin{figure}
\begin{center}
\includegraphics{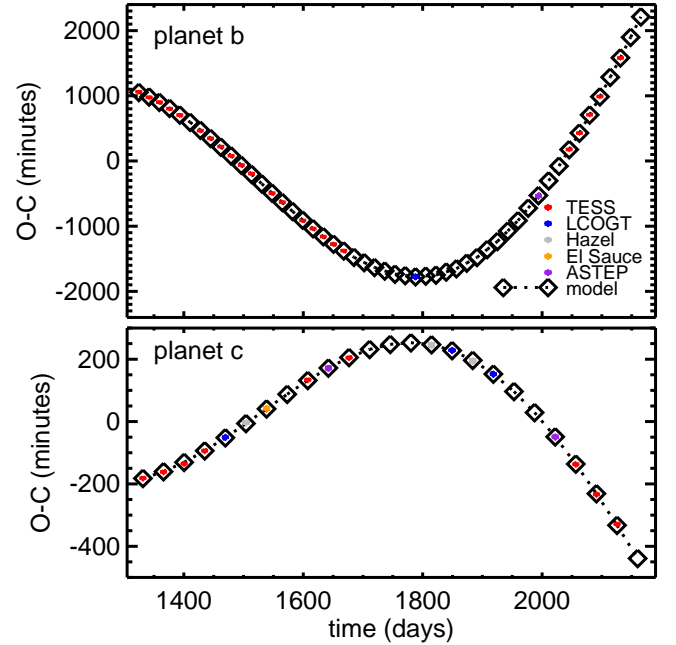}
\caption{
 \label{fig:oc} 
Observed mid-transit times (diamonds) with subtracted linear ephemeris for TOI-216\inn (top) and TOI-216\out (bottom), with the best fit model overplotted (diamonds, dotted line).
}
\end{center}
\end{figure}

TOI-216b and c exhibit transit timing variations -- plotted in Fig. \ref{fig:oc} -- due to the near-resonant effect (e.g., \citealt{agol05,lith12}). We have not yet observed a full super-period (which depends on the planets' proximity to the 2:1 resonance). The amplitude depends on the perturbing planet's mass and the free eccentricity of the transiting and perturbing planets. We previously found a significant free eccentricity, with the exact value and partition between planets degenerate with planet mass \citep{daws19}.

We fit the transit times using our N-body integrator TTV  model \citep{daws14}. Our model contains five parameters for each planet: the mass $M$, orbital period $P$, mean longitude $\lambda$, eccentricity $e$, and argument of periapse $\omega$. All orbital elements are osculating orbital elements at the epoch 1325.3279 days. For each planet, we fix the impact parameter $b$ to the value in Table~\ref{tab:216lc} and set the longitude of ascending node on the sky to $\Omega_{\rm sky}=0$. We use the conventional coordinate system where the $X-Y$ plane is the sky plane and the $Z$ axis points toward the observer. We will explore other possibilities for $b$ and $\Omega_{\rm sky}$ in Section \ref{subsec:joint}.

\begin{figure}
\begin{center}
\includegraphics{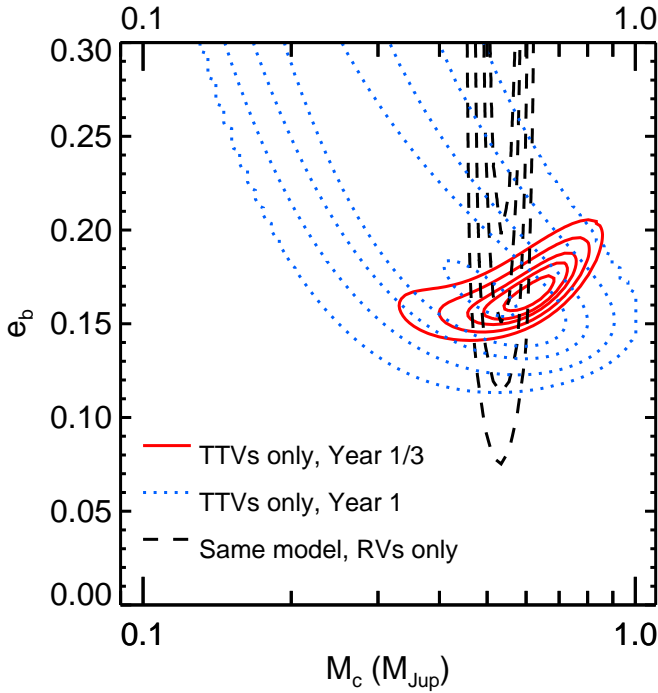}
\caption{ 
 \label{fig:contour} 
Contours of $\chi^2$ for the fit to transit times only. With ground-based transits and Year 1 of TESS data (blue-dashed), fits to the transit times only result in degeneracy between the inner planet's (osculating) eccentricity and the outer planet's mass. The levels are $\chi^2$ =  42, 47, 52, 62, and 77 and the best-fit solutions occupy the innermost contour. The black contours show $\chi^2$ for these same solutions compared to the RV measurements, with an RV offset for each instrument as the only free parameter. The levels are  $\chi^2 = $540, 560, 600, and 650 and the best-fit solutions occupy the innermost contour. The RV measurements break the degeneracy between mass and eccentricity. The addition of Year 3 TESS Extended Mission data (through sector 30; red solid) reduces the degeneracy between mass and eccentricity and shows good agreement with the RV data. The levels are  $\chi^2$ =  65, 70, 75, 85, and 100 and the best-fit solutions occupy the innermost contour.}
\end{center}
\end{figure}

We begin by fitting the transit times only. To explore the degeneracy between mass and eccentricity, we use the Levenberg-Marquardt alogrithm implemented in IDL {\tt mpfit} \citep{mark09} to minimize the $\chi^2$ on a grid of $(M_\out, e_\inn)$. { We first use a dataset consisting of all the ground-based data and the TESS Year 1 data. We contour the total $\chi^2$ for thirty-four transit times and ten free parameters, i.e., twenty-four degrees of freedom in Fig.~\ref{fig:contour}. Even with the additional transits since \citet{daws19}, the fits still suffer from a degeneracy between mass and eccentricity, though the relationship is much tighter.} We integrate a random sample of 100 solutions with $\chi^2 < 50$ and find that they all librate in the 2:1 resonance with the resonant angle involving the longitude of periapse of the inner planet. { Next we add TESS Year 3 transits from the TESS Extended Mission, which became available while this manuscript was under review. This very extended baseline for the TTVs substantially reduces the degeneracy between mass and eccentricity. We also contour the total $\chi^2$ for these forty-three transit times and ten free parameters, i.e., thirty-three degrees of freedom in Fig.~\ref{fig:contour}.}

\begin{figure}
\begin{center}
\includegraphics{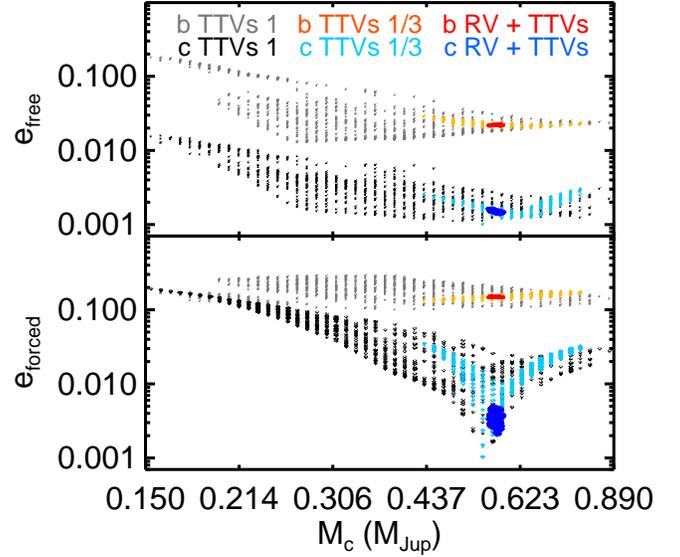}
\caption{
 \label{fig:long}  
Long-term ($10^6$~days) behavior of solutions with $\chi^2<60$ (TTV-only fits with ground-based transits and Year 1 of TESS data, gray and and black, which have discrete values because they are performed on a grid);  $\chi^2<80$ (TTV-only fits with addition of Year 3 TESS Extended Mission data, orange and light blue); and from the MCMC posterior for the joint RV-TTV fit (red and blue, Table \ref{tab:216}). Row 1: $e_{\rm free}$ (approximated as $(e_{\rm max}-e_{\rm min})/2$); Row 2: $e_{\rm forced}$ (approximated as  $(e_{\rm max}+e_{\rm min})/2$).
}
\end{center}
\end{figure}

A resonant or near-resonant planet's total eccentricity is the vector sum of its free and forced eccentricity. The forced eccentricity vector is dictated by the resonant dynamics, and the free eccentricity vector oscillates about the tip of the forced vector. Dissipation (e.g., eccentricity damping from the disk) tends to damp the free eccentricity, and other perturbations (e.g., from a third planet) can excite it. We estimate the eccentricity components from simulations as $e_{\rm free} = (e_{\rm max}-e_{\rm min})/2$) and $e_{\rm forced} = (e_{\rm max}+e_{\rm min})/2$). The transit times continue to provide evidence for a moderate free eccentricity (Fig. \ref{fig:long}), but pinning down the outer planet's mass with a joint radial-velocity fit will allow for a { tighter} constraint.  

\subsection{Joint transit and radial velocity fit}
\label{subsec:joint}

We perform a joint N-body fit to the transit times and radial velocities, imposing two additional constraints. One constraint is the transit exclusion intervals reported in \citet{daws19}. The other is the light curve joint posterior of stellar density $\rhocirc$ vs. impact parameter $b$ (Table \ref{tab:216lc}; Eqn. \ref{eqn:rhocirc}) combined with the $\rho_\star$ posterior from \citet{daws19}. Following \citet{daws12}, we convert the $\rhocirc$ vs. $b$ to a $g$ vs. $b$ posterior according to $g = \left(\rhocirc/\rho_\star\right)^{1/3}$. We use the $g$ vs. $b$ posterior to add a term to the likelihood based on $e$ and $\omega$ using $g = (1 + e \sin \omega)/(1-e^2)^{1/2}$. We compute the sky-plane inclination for the dynamical model from $b$ according to $ \sin i_{\rm sky}  a/R_\star= (1 + e \sin \omega)/(1-e^2)$.

Following \citet{daws14}, we derive posteriors for the parameters using Markov Chain Monte Carlo with the Metropolis-Hastings algorithm. Instead of including the orbital period and mean longitude at epoch 1325.3279 days as parameters in the MCMC, we optimize them at each jump (i.e., each MCMC step) using the Levenberg-Marquardt algorithm (i.e., optimizing the dynamical model). We also optimize the radial velocity offset for each instrument at each jump.  We fit for a radial velocity jitter term for each of the three instruments; even though we identified possible periodicities in Fig. \ref{fig:resid}, none are strong enough to justify to explicitly modeling (e.g., with false alarm probabilities of 0.2, 0.002, and 0.02 for the highest peak in the periodogram for PFS, FEROS, and HARPS respectively; \citealt{cumm04}).  We visually inspect each parameter for convergence. We know there is a mutual inclination perpendicular to the sky plane because the inner planet exhibits grazing transits and the outer planet does not, To also allow for a mutual inclination parallel to the sky plane, we fit for $\Delta \Omega_{\rm sky}$, the difference in longitude of ascending node. 

Following \citet{daws19}, we perform two fits with different priors to assess the sensitivity to these priors.  The first solution (Table~\ref{tab:216}) imposes uniform priors on eccentricities and log uniform priors on mass (i.e., priors that are uniform in log space). The second imposes uniform priors on mass and log-uniform priors on eccentricity. All other fitted parameters (orbital period, mean longitude, argument of periapse, radial velocity jitters, difference in longitude of ascending node) have uniform priors. With the dataset in \citet{daws19}, the results were very sensitive to the priors; with the new dataset that includes radial-velocities and an expanded TTV baseline, the results with different priors are nearly indistinguishable. { In both cases,} we impose the three-sigma limits on change in transit duration derived in Section \ref{sec:lc}. { We measure a small but significant mutual inclination of 1.2--3.9$^\circ$ (95\% confidence interval)}.

\begin{deluxetable}{rrl}
 \tablecaption{ Planet Parameters (Osculating Orbital Elements at Epoch 1325.3279 days) for \thisstarinn and \thisstarout Derived from joint TTV/RV fit \label{tab:216} }
 \tablehead{
 \colhead{Parameter}    & \colhead{Value\tablenotemark{a}}}
 \startdata
 $M_\star (M_\odot)$\tablenotemark{b} & 0.77\\
  $R_\star (R_\odot)$\tablenotemark{b} & 0.748\\\\
$M_\inn$ ($M_{\rm Jup}$) & 0.059 &$^{+0.002}_{-0.002}$ \\
$P_\inn$ & 17.0968 & $^{+0.0007}_{-0.0007}$ \\
$e_\inn$ & 0.160 & $^{+0.003}_{-0.002}$ \\
$\varpi_\inn$ (deg.) & 291.8&$^{+0.8}_{-1.0}$\\
$\lambda_\inn$ (deg) & 82.5&$^{+0.2}_{-0.3}$\\
$\Omega_{\inn, \rm sky}$ (deg) & 0 \\
$i_{\inn,\rm sky}$ (deg) & 88.60&$^{+0.03}_{-0.04}$\\\\
$M_\out$ ($M_{\rm Jup}$) & 0.56 &$^{+0.02}_{-0.02}$\\
$P_\out$ & 34.5516& $^{+0.0003}_{-0.0003}$ \\
$e_\out$ & 0.0046& $^{+0.0027}_{-0.0012}$ \\
$\varpi_\out$ (deg.) &190&$^{+30}_{-50}$\\
$\lambda_\out$ (deg) & 27.8$^{+1.7}_{-1.5}$\\
$\Delta \Omega_{\rm sky} = \Omega_{\out, \rm sky} $ (deg) & -1&$^{+2}_{-2}$\\
$i_{\inn,\rm sky}$ (deg) & 89.84&$^{+0.10}_{-0.08}$\\\\
\\
$2\lambda_\out - \lambda_\inn -\varpi_\inn$ (deg).&42.5&$^{+0.4}_{-0.3}$\\
$i_{\rm mut}$(deg)\tablenotemark{c} & 2.0&$^{+1.2}_{-0.5}$\\
\\
Jitter (m/s):\\
HARPS & 8&$^{+3}_{-2}$\\
FEROS & 22&$^{+5}_{-4}$\\
PFS & 7.7&$^{+1.7}_{-1.3}$\\
\enddata
 \tablenotetext{a}{As a summary statistic we report the median and 68.3\% confidence interval of the posterior distribution. An example fit with high precision suitable for numerical integration is given in Table \ref{tab:ex}.}
 \tablenotetext{b}{Stellar parameters fixed to the values reported by \citet{daws19}: 0.77$\pm0.03 M_\odot$ and 0.748$\pm0.015 R_\odot$. Uncertainties in estimated planetary masses only account for the dynamical fitting, i.e., they \emph{do not} include uncertainties in the star's mass.}
 \tablenotetext{c}{95\% confidence interval is 1.2--3.9$^\circ$. { The 99.7\% confidence interval is 1.1--4.3$^\circ$.}}
\end{deluxetable}

To ensure that our results are robust and that the parameter space has been thoroughly explored by the fitting algorithm (particularly the degeneracy between eccentricity and mass), we carry out a fit using a different N-body code and fitting algorithm. We use the Python Tool for Transit Variations (\texttt{PyTTV}; \citet{Korth2020})
to fit the transit times (Table \ref{tab:216lc}), RVs (Table \ref{tab:rvs}), and the stellar parameters reported in \citet{daws20}. The parameter estimation is carried out by a joint-N-body fit using \texttt{Rebound} with the IAS15 integrator \citep{rein12, rein15} and \texttt{Reboundx} \citep{tama20}, to model all the observables without approximations. Within the simulation, carried out in barycentric coordinates, a common coordinate system was chosen where the $x-y$ plane is the plane of sky. The $x$-axis points to the East, the $y$-axis points to the North, and the $z$-axis points to the observer. $\Omega$ is measured from East to North, and $\omega$ is measured from the plane of sky. For the parameter estimation with \texttt{PyTTV}, the gravitational forces between the planets and the influence of general relativity (GR) were considered, in case the influence of GR is significant enough to be visible in the TTVs; { for this system, general relativistic precession does not significantly affect the TTVs}. The parameter estimation is initialized using Rayleigh priors on eccentricities \citep{vane19} and uniform priors on log values for the planetary masses. The estimation of physical quantities from the TTV signal is done in two steps. First, the posterior mode is found using the Differential Evolution algorithm implemented in \texttt{PyTransit} \citep{parv15}. The optimization is carried out varying the planetary masses, orbital periods, inclinations, eccentricities and arguments of periastron. The eccentricity and argument of periastron are mapped from sampling-parameters $\sqrt{e}\cos\omega$ and  $\sqrt{e}\sin\omega$. The longitudes of the ascending nodes are fixed for both planets. After the posterior mode is found, a sample from the posterior is obtained using the affine-invariant Monte Carlo Markov Chain ensemble sampler \texttt{emcee} \citep{fore13}. The parameters and uncertainties are consistent with those reported in Table \ref{tab:216}.

\subsection{Dynamics and origin}

\begin{figure}
\begin{center}
\includegraphics{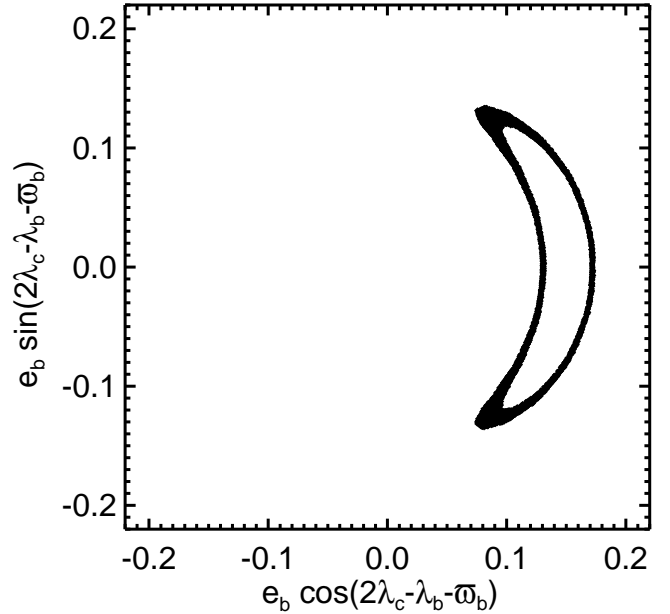}
\caption{
 \label{fig:surf} 
Example trajectory for TOI-216 solution. The trajectory does not pass through the origin, indicating libration about a fixed point. The offset from the origin is the forced eccentricity.
}
\end{center}
\end{figure}

We randomly draw 1000 solutions from the posterior for longer integrations of $10^6$ days using {\tt mercury6} \citep{cham96}. We compute the libration amplitudes for the  2:1 resonant angle $2\lambda_\out - \lambda_\inn -\varpi_\inn$, where the longitude of periapse $\varpi_\inn = \omega_\inn + \Omega_\inn$. We can now  definitively determine that the system is librating in resonance. The libration amplitude is well-constrained to $60^{+2}_{-2}$ degrees. The very small uncertainty in the libration amplitude leads us to believe that its moderate value is real, not an artifact of imperfect characterization (e.g., \citealt{mill18}). Fig. \ref{fig:surf} shows a trajectory demonstrating libration about a fixed point. In our solutions, the other 2:1 resonant angle $2\lambda_\out - \lambda_\inn -\varpi_\out$ always circulates. However, the outer planet's eccentricity and longitude of periapse are poorly constrained, so it is possible that solutions where this angle librates are also consistent with the data. The inner planet has a forced  eccentricity (Fig. \ref{fig:long}) of $0.146^{+0.003}_{-0.003}$ and a modest but significant free eccentricity of $0.0222^{+0.0005}_{-0.0003}$. The outer planet has very small forced and free eccentricities of $0.0048^{+0.0028}_{-0.0013}$ and $0.00162^{+0.00011}_{-0.00010}$ respectively. In our integrated solutions, we find that the timescale for the biggest variations in both planets' semi-major axes and the inner planet's eccentricity is the resonant libration timescale (approximately 5 years), and the outer planet's eccentricity varies on apsidal alignment timescale (approximately 25 years). This system falls into the regime of resonant TTVs (e.g., \citealt{nesv16}) -- rather than the more commonly characterized near-resonant TTVs (e.g., \citealt{lith12})-- and we expect the TTVs to oscillate on the libration timescale, which has not yet been fully covered by the observational baseline (Fig. \ref{fig:oc}). We integrate a random subset of 200 solutions for 1 Myr and find that all are stable.

\begin{figure*}
\begin{center}
\includegraphics{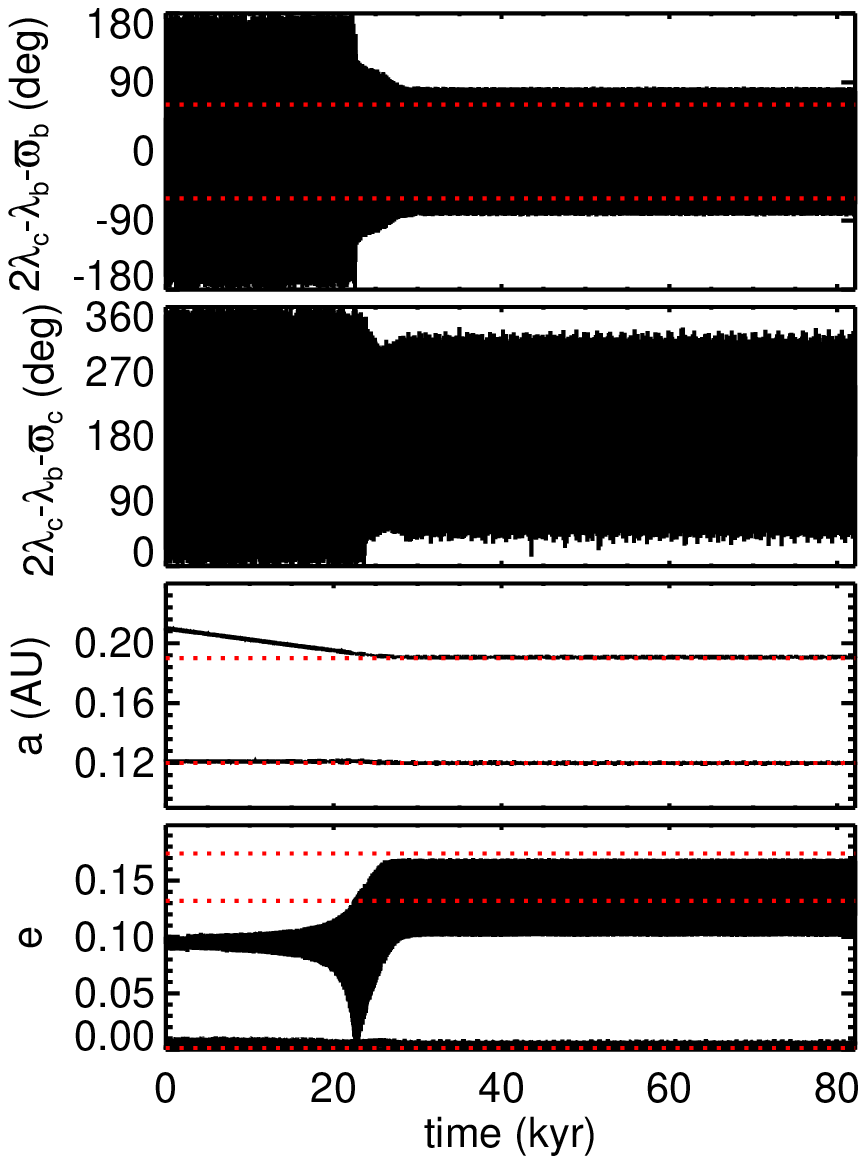}
\includegraphics{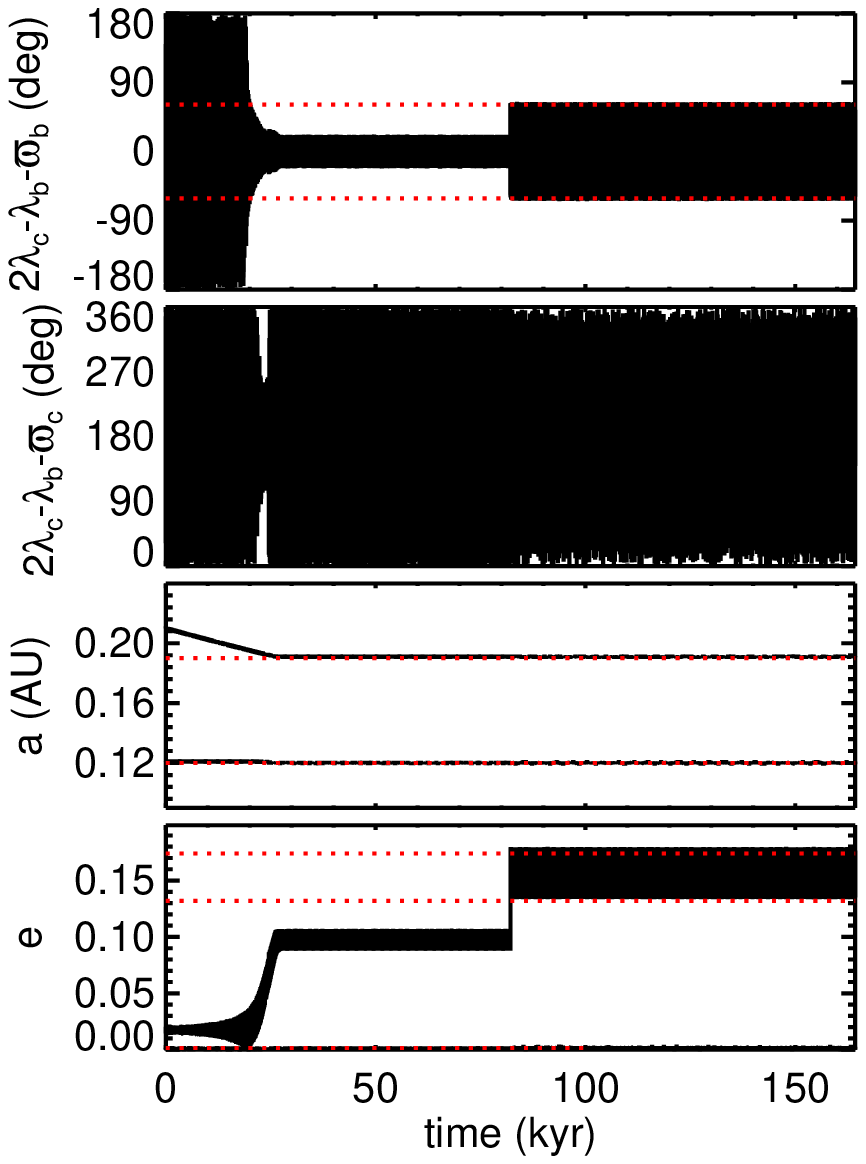}
\caption{
 \label{fig:migr} 
Examples of resonance capture through short distance convergent migration of planet c. Rows 1--2: resonant angle; row 3: semi-major axis of planet c (top black line) and planet b (bottom black line); row 4: eccentricity of planet b (top black line) and planet b (bottom black line). Values from simulations with the present-day observed orbits (i.e., range of oscillation) are marked with a dotted red line. 
}
\end{center}
\end{figure*}

One possible scenario for establishing the resonance is convergent migration of planet c preceded or followed by perturbations of planet b by another body (Fig. \ref{fig:migr}). We show example simulations that apply a migration force using the user-defined force feature of {\tt mercury6} (as described in \citealt{wolf12}); the simulations are done in 3D with initial inclinations set to the present day values. Planet c migrates a short distance (0.8\% of its initial semi-major axis) toward planet b and captures b into resonance. In the first example (top), planet b starts with a modestly eccentric (bottom) orbit ($e=0.0798$). In the second example, planet b starts with a lower eccentricity ($e=0.02$) and is captured into resonance with a tight libration amplitude. At about 80,000 years, planet b's orbit is disturbed, which we approximate as an instantaneous change in the magnitude and direction of its eccentricity vector. The first history results in large amplitude libration of the resonant angle involving $\varpi_c$ and the second in circulation; because we are not confident that only circulating resonant angles are compatible with the data, we cannot use this distinction to favor one history over the other.

These example dynamical histories are compatible with in situ formation, but are potentially compatible with long distance migration as well. In the in situ formation scenario, three or more planets form in situ, and planet c migrates a tiny distance toward b. A third planet jostles b -- exciting its eccentricity and mutual inclination -- before or after c's migration. In the long distance migration scenario, the disturbing third planet could have migrated earlier. As future work, orbital dynamics simulations could place limits on the properties of this possible third planet. Although we have invoked a third planet in the example scenarios, we have not ruled out the possibility that planet b itself disturbed planet c, in a process separate from migration.

\begin{figure}
\begin{center}
\includegraphics{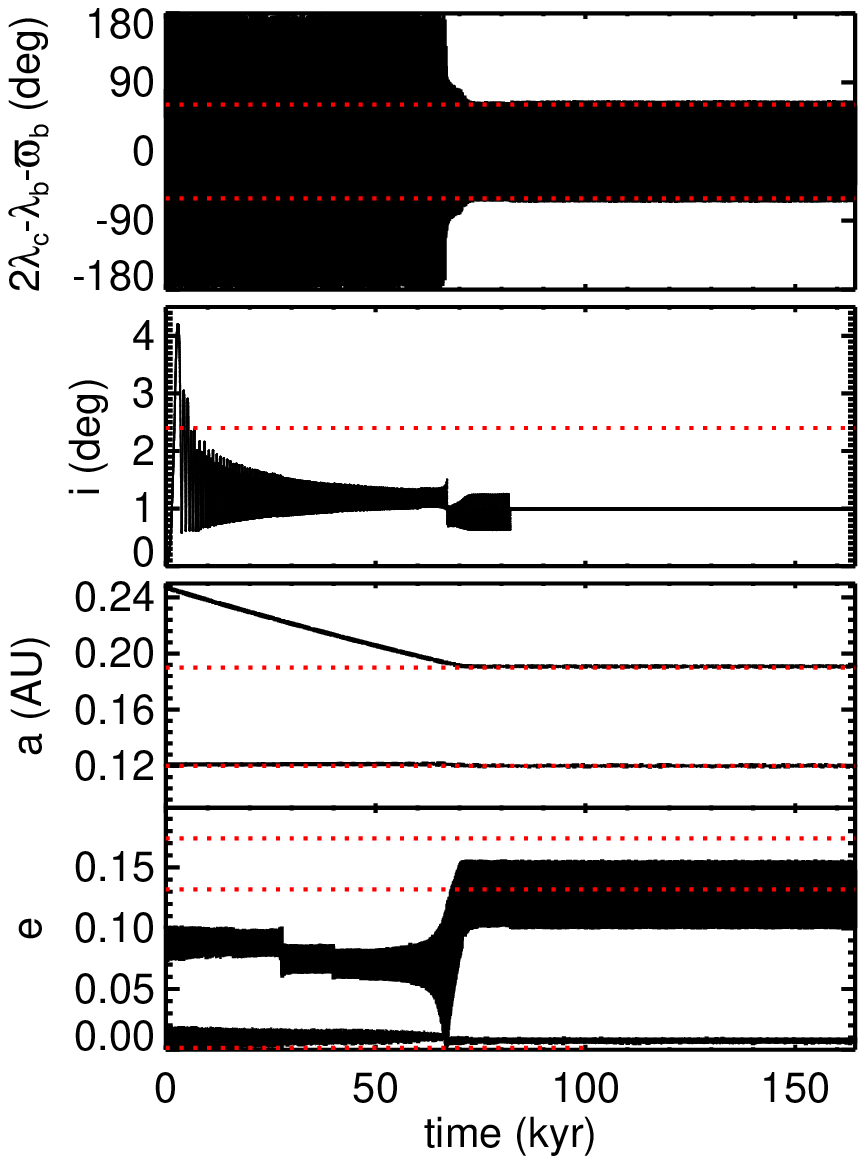}
\caption{
 \label{fig:ires} 
 Examples of excitation of a mutual inclination during short distance convergent migration of planet c. Row 1: eccentricity resonant angle; row 2: mutual inclination; row 3: semi-major axis of planet c (top black line) and planet b (bottom black line); row 4: eccentricity of planet b (top black line) and planet b (bottom black line). Values from simulations with the present-day observed orbits (i.e., range of oscillation) are marked with a dotted red line. 
}
\end{center}
\end{figure}

{ Another hypothesis for the mutual inclination is that it was excited when the planets passed through the 4:2 inclination resonance during migration \citep{tho03}. Precession from the proto-planetary disk can separate the 4:2 inclination resonances from the 2:1 eccentricity resonance. Figure \ref{fig:ires} shows a proof of concept simulation where the potential of the proto-planetary disk is approximated as a $J_2$ oblateness term for the stellar potential. A small but significant mutual inclination is excited.}

\begin{figure}
\begin{center}
\includegraphics{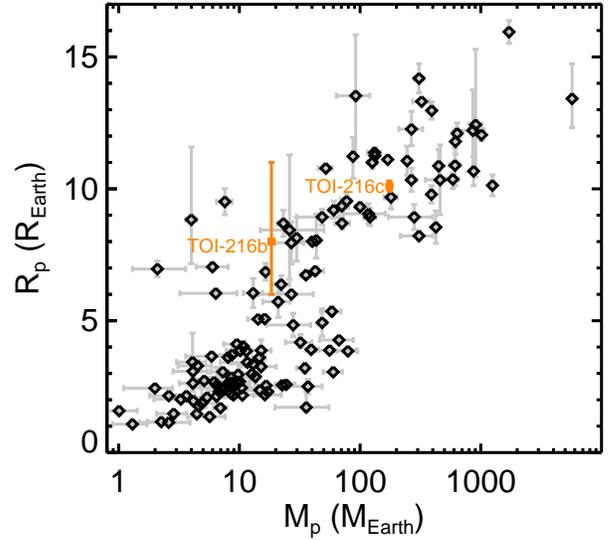}
\caption{\label{fig:rm} 
Warm (10--200~day orbital period) planets with both mass and radius measurements (exoplanet.eu; \citealt{schn11}), including TOI-216 (orange). 
}
\end{center}
\end{figure}

We place planet b and c on a mass-radius plot of warm exoplanets in Fig.~\ref{fig:rm}. TOI-216c has a typical radius for its mass; its bulk density is 0.885$^{+0.014}_{-0.13}$ gcm$^{-3}$. TOI-216b likely also has a typical radius for its mass, but because of its grazing transit (Section \ref{sec:lc}) and the resulting degeneracy with impact parameter, we cannot rule out a large radius that would result in a low density for its mass compared to similar mass planets. Its poorly-constrained bulk density is 0.17$^{+0.18}_{-0.10}$ gcm$^{-3}$.

\section{Conclusion}
\label{sec:discuss}
TOI-216b and c are now a very precisely characterized (with the exception of planet b's radius) pair of warm, large exoplanets. 
Radial velocity measurements using HARPS, FEROS, and PFS broke a degeneracy between mass and eccentricity in the TTV-only fits { that was particularly severe before the TESS Extended Mission observations}, and an expanded TTV baseline from \TESS and an ongoing ground-based transit observing campaign increased the precision of the fits. We can now better assess the consistency of its properties with different theories for formation and evolution of giant planets orbiting close to their stars.

We now know that TOI-216c is a warm Jupiter ($0.54 \pm 0.02$ Jupiter masses) with a mass and radius typical of other 10--200 day giant planets (Fig. \ref{fig:rm}). TOI-216b has a mass similar to Neptune's (18.4$\pm0.6 M_\oplus$). Its radius is not well constrained due to its grazing transits; its radius may very well be typical for its mass (Fig. \ref{fig:rm}). { Given the large uncertainty in planet b's radius}, no inflation mechanisms or scenarios requiring formation beyond the ice line (e.g., \citealt{lee16}) are required to explain { either planet's radius.}

Furthermore, we know now that TOI-216b and c are not just near but librating in the 2:1 resonance. The argument involving the longitude of periapse of TOI-216b librates with an amplitude of $63^{+3}_{-3}$ degrees. TOI-216b has a small but significant free eccentricity $0.0222^{+0.0005}_{-0.0003}$. The mutual inclination with respect to TOI-216c is between 1.2--3.9$^\circ$ (95\% confidence interval). The libration amplitude, free eccentricity, and mutual inclination imply a disturbance of TOI-216b before or after resonance capture, perhaps by an undetected third planet.  The orbital properties can be consistent either with in situ formation, with resonance capture through a very short distance migration, or long distance migration. Future origins scenarios must match these precisely constrained properties. { Future atmospheric characterization by the James Webb Space Telescope (JWST) may help distinguish between these origin scenarios. If the planet formed outside the water snow line, we expect its C/O ratio to be significantly smaller than that of its host star; in-situ formation, on the other hand, would imply C/O ratios closer to the star \citep{espi17}. Simulations performed with PandExo \citep{pandexo} show that  water features in the spectra that constrain the C/O ratio may be detectable for TOI-216c with the JWST Near Infrared Imager and  Slitless Spectrograph (NIRISS), even beneath a moderate cloud layer.}

This system will benefit from continued long-term radial-velocity and transit monitoring. Long-term radial velocity monitoring could reveal the presence of additional planets in the system, though stellar activity poses a challenge for detecting low amplitude signals. Observations by the \TESS Extended Mission are { continuing} and will { further} increase the baseline of observations. The new \TESS observations may allow us to { better constrain the change in impact parameter tentatively detected here for TOI-216b}, allowing for tighter constraints on the mutual inclination. 

\acknowledgments
 We thank the \TESS Mission team and follow up working group for the valuable dataset. We acknowledge the use of public \TESS Alert data from pipelines at the \TESS Science Office and at the \TESS Science Processing Operations Center.  This paper includes data collected by the \TESS mission, which are publicly available from the Mikulski Archive for Space Telescopes (MAST).  Resources supporting this work were provided by the NASA High-End Computing (HEC) Program through the NASA Advanced Supercomputing (NAS) Division at Ames Research Center for the production of the SPOC data products.
 
This research has made use of the Exoplanet Follow-up Observation Program website, which is operated by the California Institute of Technology, under contract with the National Aeronautics and Space Administration under the Exoplanet Exploration Program. This work has made use of observations from the Las Cumbres Observatory network.  This work has made use of data from the European Space Agency (ESA) mission {\it Gaia} (\url{https://www.cosmos.esa.int/gaia}), processed by the {\it Gaia} Data Processing and Analysis Consortium (DPAC, \url{https://www.cosmos.esa.int/web/gaia/dpac/consortium}). Funding for the DPAC has been provided by national institutions, in particular the institutions participating in the {\it Gaia} Multilateral Agreement. We acknowledge support for ASTEP from the French and Italian Polar Agencies, IPEV and PNRA, and from Université Côte d’Azur under Idex UCAJEDI (ANR-15-IDEX-01). This paper includes data gathered with the 6.5 meter Magellan Telescopes located at Las Campanas Observatory, Chile.
 
 RID and JD gratefully acknowledge support by NASA XRP NNX16AB50G, NASA XRP 80NSSC18K0355, NASA \TESS GO 80NSSC18K1695, and the Alfred P. Sloan Foundation's Sloan Research Fellowship. The Center for Exoplanets and Habitable Worlds is supported by the Pennsylvania State University, the Eberly College of Science, and the Pennsylvania Space Grant Consortium.  JD gratefully acknowledges support and hospitality from the pre-doctoral program at the Center for Computational Astrophysics, Flatiron Institute. Research at the Flatiron Institute is supported by the Simons Foundation. This research made use of computing facilities from Penn State's Institute for CyberScience Advanced CyberInfrastructure. 
 
 We thank Caleb Ca\~nas and David Nesvorn{\'y} for helpful comments and discussions. { We thank the referee for a helpful report.}
 
R.B.\ acknowledges support from FONDECYT Project 11200751 and from CORFO project N$^\circ$14ENI2-26865. A.J.\, R.B.\, and M.H.\ acknowledge support from project IC120009 ``Millennium Institute of Astrophysics (MAS)'' of the Millenium Science Initiative, Chilean Ministry of Economy. A.J.\ acknowledges additional support from FONDECYT project 1171208. This research received funding from the European Research Council (ERC) under the European Union's Horizon 2020 research and innovation programme (grant agreement n$^\circ$ 803193/BEBOP), and from the Science and Technology Facilities Council (STFC; grant n$^\circ$ ST/S00193X/1). J.~Korth acknowledges support by DFG grants PA525/19-1 within the DFG Schwerpunkt SPP 1992, Exploring the Diversity of Extrasolar Planets. Part of this research was carried out at the Jet Propulsion Laboratory, California Institute of Technology, under a contract with the National Aeronautics and Space Administration (NASA).

 \software{ mpfit \citep{mark09}, TAP\citep{gaza12}, Tapir \citep{Jensen:2013}, \TESS pipeline \citep{jenk16,twic18,li19}, AstroImageJ \citep{Collins:2017}, \texttt{exoplanet} \citep{exoplanet:exoplanet}, \texttt{astropy} \citep{exoplanet:astropy13, exoplanet:astropy18}, \texttt{celerite} \citep{exoplanet:foremanmackey17, exoplanet:foremanmackey18}, \texttt{starry} \citep{exoplanet:luger18}, \texttt{pymc3} \citep{exoplanet:pymc3}, {\tt theano} \citep{exoplanet:theano},
 \texttt{ceres} \citep{ceres}, \texttt{PyTTV}, \texttt{PyTransit} \citep{parv15}, \texttt{emcee} \citep{fore13}, {\tt eleanor} \citep{fein19}}
 
\newpage
\bibliographystyle{aasjournal}
\bibliography{biblio}

\begin{thebibliography}{}
\expandafter\ifx\csname natexlab\endcsname\relax\def\natexlab#1{#1}\fi
\providecommand{\url}[1]{\href{#1}{#1}}
\providecommand{\dodoi}[1]{doi:~\href{http://doi.org/#1}{\nolinkurl{#1}}}
\providecommand{\doeprint}[1]{\href{http://ascl.net/#1}{\nolinkurl{http://ascl.net/#1}}}
\providecommand{\doarXiv}[1]{\href{https://arxiv.org/abs/#1}{\nolinkurl{https://arxiv.org/abs/#1}}}

\bibitem[{{Agol} {et~al.}(2005){Agol}, {Steffen}, {Sari}, \&
  {Clarkson}}]{agol05}
{Agol}, E., {Steffen}, J., {Sari}, R., \& {Clarkson}, W. 2005, \mnras, 359,
  567, \dodoi{10.1111/j.1365-2966.2005.08922.x}

\bibitem[{{Anderson} {et~al.}(2020){Anderson}, {Lai}, \& {Pu}}]{ande20}
{Anderson}, K.~R., {Lai}, D., \& {Pu}, B. 2020, \mnras, 491, 1369,
  \dodoi{10.1093/mnras/stz3119}

\bibitem[{{Anglada-Escud{\'e}} {et~al.}(2010){Anglada-Escud{\'e}},
  {L{\'o}pez-Morales}, \& {Chambers}}]{angl10}
{Anglada-Escud{\'e}}, G., {L{\'o}pez-Morales}, M., \& {Chambers}, J.~E. 2010,
  \apj, 709, 168, \dodoi{10.1088/0004-637X/709/1/168}

\bibitem[{{Astropy Collaboration} {et~al.}(2013){Astropy Collaboration},
  {Robitaille}, {Tollerud}, {Greenfield}, {Droettboom}, {Bray}, {Aldcroft},
  {Davis}, {Ginsburg}, {Price-Whelan}, {Kerzendorf}, {Conley}, {Crighton},
  {Barbary}, {Muna}, {Ferguson}, {Grollier}, {Parikh}, {Nair}, {Unther},
  {Deil}, {Woillez}, {Conseil}, {Kramer}, {Turner}, {Singer}, {Fox}, {Weaver},
  {Zabalza}, {Edwards}, {Azalee Bostroem}, {Burke}, {Casey}, {Crawford},
  {Dencheva}, {Ely}, {Jenness}, {Labrie}, {Lim}, {Pierfederici}, {Pontzen},
  {Ptak}, {Refsdal}, {Servillat}, \& {Streicher}}]{exoplanet:astropy13}
{Astropy Collaboration}, {Robitaille}, T.~P., {Tollerud}, E.~J., {et~al.} 2013,
  \aap, 558, A33, \dodoi{10.1051/0004-6361/201322068}

\bibitem[{{Astropy Collaboration} {et~al.}(2018){Astropy Collaboration},
  {Price-Whelan}, {Sip{\H o}cz}, {G{\"u}nther}, {Lim}, {Crawford}, {Conseil},
  {Shupe}, {Craig}, {Dencheva}, {Ginsburg}, {VanderPlas}, {Bradley},
  {P{\'e}rez-Su{\'a}rez}, {de Val-Borro}, {Aldcroft}, {Cruz}, {Robitaille},
  {Tollerud}, {Ardelean}, {Babej}, {Bach}, {Bachetti}, {Bakanov}, {Bamford},
  {Barentsen}, {Barmby}, {Baumbach}, {Berry}, {Biscani}, {Boquien}, {Bostroem},
  {Bouma}, {Brammer}, {Bray}, {Breytenbach}, {Buddelmeijer}, {Burke},
  {Calderone}, {Cano Rodr{\'{\i}}guez}, {Cara}, {Cardoso}, {Cheedella},
  {Copin}, {Corrales}, {Crichton}, {D'Avella}, {Deil}, {Depagne}, {Dietrich},
  {Donath}, {Droettboom}, {Earl}, {Erben}, {Fabbro}, {Ferreira}, {Finethy},
  {Fox}, {Garrison}, {Gibbons}, {Goldstein}, {Gommers}, {Greco}, {Greenfield},
  {Groener}, {Grollier}, {Hagen}, {Hirst}, {Homeier}, {Horton}, {Hosseinzadeh},
  {Hu}, {Hunkeler}, {Ivezi{\'c}}, {Jain}, {Jenness}, {Kanarek}, {Kendrew},
  {Kern}, {Kerzendorf}, {Khvalko}, {King}, {Kirkby}, {Kulkarni}, {Kumar},
  {Lee}, {Lenz}, {Littlefair}, {Ma}, {Macleod}, {Mastropietro}, {McCully},
  {Montagnac}, {Morris}, {Mueller}, {Mumford}, {Muna}, {Murphy}, {Nelson},
  {Nguyen}, {Ninan}, {N{\"o}the}, {Ogaz}, {Oh}, {Parejko}, {Parley}, {Pascual},
  {Patil}, {Patil}, {Plunkett}, {Prochaska}, {Rastogi}, {Reddy Janga},
  {Sabater}, {Sakurikar}, {Seifert}, {Sherbert}, {Sherwood-Taylor}, {Shih},
  {Sick}, {Silbiger}, {Singanamalla}, {Singer}, {Sladen}, {Sooley},
  {Sornarajah}, {Streicher}, {Teuben}, {Thomas}, {Tremblay}, {Turner},
  {Terr{\'o}n}, {van Kerkwijk}, {de la Vega}, {Watkins}, {Weaver}, {Whitmore},
  {Woillez}, {Zabalza}, \& {Astropy Contributors}}]{exoplanet:astropy18}
{Astropy Collaboration}, {Price-Whelan}, A.~M., {Sip{\H o}cz}, B.~M., {et~al.}
  2018, \aj, 156, 123, \dodoi{10.3847/1538-3881/aabc4f}

\bibitem[{{Batalha} {et~al.}(2017){Batalha}, {Mandell}, {Pontoppidan},
  {Stevenson}, {Lewis}, {Kalirai}, {Earl}, {Greene}, {Albert}, \&
  {Nielsen}}]{pandexo}
{Batalha}, N.~E., {Mandell}, A., {Pontoppidan}, K., {et~al.} 2017, \pasp, 129,
  064501, \dodoi{10.1088/1538-3873/aa65b0}

\bibitem[{{Boisse} {et~al.}(2009){Boisse}, {Moutou}, {Vidal-Madjar}, {Bouchy},
  {Pont}, {H{\'e}brard}, {Bonfils}, {Croll}, {Delfosse}, {Desort}, {Forveille},
  {Lagrange}, {Loeillet}, {Lovis}, {Matthews}, {Mayor}, {Pepe}, {Perrier},
  {Queloz}, {Rowe}, {Santos}, {S{\'e}gransan}, \& {Udry}}]{bois09}
{Boisse}, I., {Moutou}, C., {Vidal-Madjar}, A., {et~al.} 2009, \aap, 495, 959,
  \dodoi{10.1051/0004-6361:200810648}

\bibitem[{{Brahm} {et~al.}(2017){Brahm}, {Jord{\'a}n}, \& {Espinoza}}]{ceres}
{Brahm}, R., {Jord{\'a}n}, A., \& {Espinoza}, N. 2017, \pasp, 129, 034002,
  \dodoi{10.1088/1538-3873/aa5455}

\bibitem[{{Brahm} {et~al.}(2019){Brahm}, {Espinoza}, {Jord{\'a}n}, {Henning},
  {Sarkis}, {Jones}, {D{\'\i}az}, {Jenkins}, {Vanzi}, {Zapata}, {Petrovich},
  {Kossakowski}, {Rabus}, {Rojas}, \& {Torres}}]{brahm:2019}
{Brahm}, R., {Espinoza}, N., {Jord{\'a}n}, A., {et~al.} 2019, \aj, 158, 45,
  \dodoi{10.3847/1538-3881/ab279a}

\bibitem[{{Brown} {et~al.}(2013){Brown}, {Baliber}, {Bianco}, {Bowman},
  {Burleson}, {Conway}, {Crellin}, {Depagne}, {De Vera}, {Dilday}, {Dragomir},
  {Dubberley}, {Eastman}, {Elphick}, {Falarski}, {Foale}, {Ford}, {Fulton},
  {Garza}, {Gomez}, {Graham}, {Greene}, {Haldeman}, {Hawkins}, {Haworth},
  {Haynes}, {Hidas}, {Hjelstrom}, {Howell}, {Hygelund}, {Lister}, {Lobdill},
  {Martinez}, {Mullins}, {Norbury}, {Parrent}, {Paulson}, {Petry}, {Pickles},
  {Posner}, {Rosing}, {Ross}, {Sand}, {Saunders}, {Shobbrook}, {Shporer},
  {Street}, {Thomas}, {Tsapras}, {Tufts}, {Valenti}, {Vander Horst}, {Walker},
  {White}, \& {Willis}}]{brown2013}
{Brown}, T.~M., {Baliber}, N., {Bianco}, F.~B., {et~al.} 2013, Publications of
  the Astronomical Society of the Pacific, 125, 1031, \dodoi{10.1086/673168}

\bibitem[{{Butler} {et~al.}(1996){Butler}, {Marcy}, {Williams}, {McCarthy},
  {Dosanjh}, \& {Vogt}}]{Butler1996}
{Butler}, R.~P., {Marcy}, G.~W., {Williams}, E., {et~al.} 1996, \pasp, 108,
  500, \dodoi{10.1086/133755}

\bibitem[{{Canto Martins} {et~al.}(2020){Canto Martins}, {Gomes}, {Messias},
  {de Lira}, {Le{\~a}o}, {Almeida}, {Teixeira}, {das Chagas}, {Bravo}, {Bewketu
  Belete}, \& {De Medeiros}}]{cant20}
{Canto Martins}, B.~L., {Gomes}, R.~L., {Messias}, Y.~S., {et~al.} 2020, arXiv
  e-prints, arXiv:2007.03079.
\newblock \doarXiv{2007.03079}

\bibitem[{{Carter} \& {Winn}(2009)}]{cart09}
{Carter}, J.~A., \& {Winn}, J.~N. 2009, \apj, 704, 51,
  \dodoi{10.1088/0004-637X/704/1/51}

\bibitem[{{Chambers} {et~al.}(1996){Chambers}, {Wetherill}, \& {Boss}}]{cham96}
{Chambers}, J.~E., {Wetherill}, G.~W., \& {Boss}, A.~P. 1996, \icarus, 119,
  261, \dodoi{10.1006/icar.1996.0019}

\bibitem[{{Choksi} \& {Chiang}(2020)}]{chok20}
{Choksi}, N., \& {Chiang}, E. 2020, \mnras, \dodoi{10.1093/mnras/staa1421}

\bibitem[{{Collins} {et~al.}(2017){Collins}, {Kielkopf}, {Stassun}, \&
  {Hessman}}]{Collins:2017}
{Collins}, K.~A., {Kielkopf}, J.~F., {Stassun}, K.~G., \& {Hessman}, F.~V.
  2017, \aj, 153, 77, \dodoi{10.3847/1538-3881/153/2/77}

\bibitem[{{Crane} {et~al.}(2006){Crane}, {Shectman}, \&
  {Butler}}]{craneetal2006}
{Crane}, J.~D., {Shectman}, S.~A., \& {Butler}, R.~P. 2006, Society of
  Photo-Optical Instrumentation Engineers (SPIE) Conference Series, Vol. 6269,
  {The Carnegie Planet Finder Spectrograph}, 626931, \dodoi{10.1117/12.672339}

\bibitem[{{Crane} {et~al.}(2010){Crane}, {Shectman}, {Butler}, {Thompson},
  {Birk}, {Jones}, \& {Burley}}]{Craneetal2010}
{Crane}, J.~D., {Shectman}, S.~A., {Butler}, R.~P., {et~al.} 2010, Society of
  Photo-Optical Instrumentation Engineers (SPIE) Conference Series, Vol. 7735,
  {The Carnegie Planet Finder Spectrograph: integration and commissioning},
  773553, \dodoi{10.1117/12.857792}

\bibitem[{{Crane} {et~al.}(2008){Crane}, {Shectman}, {Butler}, {Thompson}, \&
  {Burley}}]{craneetal2008}
{Crane}, J.~D., {Shectman}, S.~A., {Butler}, R.~P., {Thompson}, I.~B., \&
  {Burley}, G.~S. 2008, Society of Photo-Optical Instrumentation Engineers
  (SPIE) Conference Series, Vol. 7014, {The Carnegie Planet Finder
  Spectrograph: a status report}, 701479, \dodoi{10.1117/12.789637}

\bibitem[{{Cumming}(2004)}]{cumm04}
{Cumming}, A. 2004, \mnras, 354, 1165, \dodoi{10.1111/j.1365-2966.2004.08275.x}

\bibitem[{{Cumming} {et~al.}(1999){Cumming}, {Marcy}, \& {Butler}}]{cumm99}
{Cumming}, A., {Marcy}, G.~W., \& {Butler}, R.~P. 1999, \apj, 526, 890,
  \dodoi{10.1086/308020}

\bibitem[{{Dawson}(2020)}]{daws20}
{Dawson}, R.~I. 2020, \aj, 159, 223, \dodoi{10.3847/1538-3881/ab7fa5}

\bibitem[{{Dawson} \& {Johnson}(2012)}]{daws12}
{Dawson}, R.~I., \& {Johnson}, J.~A. 2012, \apj, 756, 122,
  \dodoi{10.1088/0004-637X/756/2/122}

\bibitem[{{Dawson} \& {Johnson}(2018)}]{daws18}
---. 2018, \araa, 56, 175, \dodoi{10.1146/annurev-astro-081817-051853}

\bibitem[{{Dawson} {et~al.}(2014){Dawson}, {Johnson}, {Fabrycky},
  {Foreman-Mackey}, {Murray-Clay}, {Buchhave}, {Cargile}, {Clubb}, {Fulton},
  {Hebb}, {Howard}, {Huber}, {Shporer}, \& {Valenti}}]{daws14}
{Dawson}, R.~I., {Johnson}, J.~A., {Fabrycky}, D.~C., {et~al.} 2014, \apj, 791,
  89, \dodoi{10.1088/0004-637X/791/2/89}

\bibitem[{{Dawson} {et~al.}(2019){Dawson}, {Huang}, {Lissauer}, {Collins},
  {Sha}, {Armstrong}, {Conti}, {Collins}, {Evans}, {Gan}, {Horne}, {Ireland},
  {Murgas}, {Myers}, {Relles}, {Sefako}, {Shporer}, {Stockdale},
  {{\v{Z}}erjal}, {Zhou}, {Ricker}, {Vand erspek}, {Latham}, {Seager}, {Winn},
  {Jenkins}, {Bouma}, {Caldwell}, {Daylan}, {Doty}, {Dynes}, {Esquerdo},
  {Rose}, {Smith}, \& {Yu}}]{daws19}
{Dawson}, R.~I., {Huang}, C.~X., {Lissauer}, J.~J., {et~al.} 2019, \aj, 158,
  65, \dodoi{10.3847/1538-3881/ab24ba}

\bibitem[{{Deck} \& {Agol}(2015)}]{deck15}
{Deck}, K.~M., \& {Agol}, E. 2015, \apj, 802, 116,
  \dodoi{10.1088/0004-637X/802/2/116}

\bibitem[{{Dong} \& {Dawson}(2016)}]{dong16}
{Dong}, R., \& {Dawson}, R. 2016, \apj, 825, 77,
  \dodoi{10.3847/0004-637X/825/1/77}

\bibitem[{{Edelson} \& {Krolik}(1988)}]{ede88}
{Edelson}, R.~A., \& {Krolik}, J.~H. 1988, \apj, 333, 646,
  \dodoi{10.1086/166773}

\bibitem[{{Espinoza} {et~al.}(2017){Espinoza}, {Fortney}, {Miguel},
  {Thorngren}, \& {Murray-Clay}}]{espi17}
{Espinoza}, N., {Fortney}, J.~J., {Miguel}, Y., {Thorngren}, D., \&
  {Murray-Clay}, R. 2017, \apjl, 838, L9, \dodoi{10.3847/2041-8213/aa65ca}

\bibitem[{{Feinstein} {et~al.}(2019){Feinstein}, {Montet}, {Foreman-Mackey},
  {Bedell}, {Saunders}, {Bean}, {Christiansen}, {Hedges}, {Luger}, {Scolnic},
  \& {Cardoso}}]{fein19}
{Feinstein}, A.~D., {Montet}, B.~T., {Foreman-Mackey}, D., {et~al.} 2019,
  \pasp, 131, 094502, \dodoi{10.1088/1538-3873/ab291c}

\bibitem[{{Foreman-Mackey}(2018)}]{exoplanet:foremanmackey18}
{Foreman-Mackey}, D. 2018, Research Notes of the American Astronomical Society,
  2, 31, \dodoi{10.3847/2515-5172/aaaf6c}

\bibitem[{{Foreman-Mackey} {et~al.}(2017){Foreman-Mackey}, {Agol},
  {Ambikasaran}, \& {Angus}}]{exoplanet:foremanmackey17}
{Foreman-Mackey}, D., {Agol}, E., {Ambikasaran}, S., \& {Angus}, R. 2017, \aj,
  154, 220, \dodoi{10.3847/1538-3881/aa9332}

\bibitem[{Foreman-Mackey {et~al.}(2019)Foreman-Mackey, Czekala, Luger, Agol,
  Barentsen, \& Barclay}]{exoplanet:exoplanet}
Foreman-Mackey, D., Czekala, I., Luger, R., {et~al.} 2019, dfm/exoplanet:
  exoplanet v0.2.1, \dodoi{10.5281/zenodo.3462740}

\bibitem[{{Foreman-Mackey} {et~al.}(2013){Foreman-Mackey}, {Hogg}, {Lang}, \&
  {Goodman}}]{fore13}
{Foreman-Mackey}, D., {Hogg}, D.~W., {Lang}, D., \& {Goodman}, J. 2013, \pasp,
  125, 306, \dodoi{10.1086/670067}

\bibitem[{{Frelikh} {et~al.}(2019){Frelikh}, {Jang}, {Murray-Clay}, \&
  {Petrovich}}]{frel19}
{Frelikh}, R., {Jang}, H., {Murray-Clay}, R.~A., \& {Petrovich}, C. 2019,
  \apjl, 884, L47, \dodoi{10.3847/2041-8213/ab4a7b}

\bibitem[{{Gazak} {et~al.}(2012){Gazak}, {Johnson}, {Tonry}, {Dragomir},
  {Eastman}, {Mann}, \& {Agol}}]{gaza12}
{Gazak}, J.~Z., {Johnson}, J.~A., {Tonry}, J., {et~al.} 2012, Advances in
  Astronomy, 2012, 697967, \dodoi{10.1155/2012/697967}

\bibitem[{{Huang} {et~al.}(2016){Huang}, {Wu}, \& {Triaud}}]{huan16}
{Huang}, C., {Wu}, Y., \& {Triaud}, A.~H.~M.~J. 2016, \apj, 825, 98,
  \dodoi{10.3847/0004-637X/825/2/98}

\bibitem[{{Jenkins} {et~al.}(2002){Jenkins}, {Caldwell}, \& {Borucki}}]{jenk02}
{Jenkins}, J.~M., {Caldwell}, D.~A., \& {Borucki}, W.~J. 2002, \apj, 564, 495,
  \dodoi{10.1086/324143}

\bibitem[{{Jenkins} {et~al.}(2010){Jenkins}, {Caldwell}, {Chandrasekaran},
  {Twicken}, {Bryson}, {Quintana}, {Clarke}, {Li}, {Allen}, {Tenenbaum}, {Wu},
  {Klaus}, {Middour}, {Cote}, {McCauliff}, {Girouard}, {Gunter}, {Wohler},
  {Sommers}, {Hall}, {Uddin}, {Wu}, {Bhavsar}, {Van Cleve}, {Pletcher},
  {Dotson}, {Haas}, {Gilliland}, {Koch}, \& {Borucki}}]{jenk10}
{Jenkins}, J.~M., {Caldwell}, D.~A., {Chandrasekaran}, H., {et~al.} 2010,
  \apjl, 713, L87, \dodoi{10.1088/2041-8205/713/2/L87}

\bibitem[{{Jenkins} {et~al.}(2016){Jenkins}, {Twicken}, {McCauliff},
  {Campbell}, {Sanderfer}, {Lung}, {Mansouri-Samani}, {Girouard}, {Tenenbaum},
  {Klaus}, {Smith}, {Caldwell}, {Chacon}, {Henze}, {Heiges}, {Latham},
  {Morgan}, {Swade}, {Rinehart}, \& {Vanderspek}}]{jenk16}
{Jenkins}, J.~M., {Twicken}, J.~D., {McCauliff}, S., {et~al.} 2016, in
  \procspie, Vol. 9913, Software and Cyberinfrastructure for Astronomy IV,
  99133E, \dodoi{10.1117/12.2233418}

\bibitem[{{Jensen}(2013)}]{Jensen:2013}
{Jensen}, E. 2013, {Tapir: A web interface for transit/eclipse observability},
  Astrophysics Source Code Library.
\newblock \doeprint{1306.007}

\bibitem[{{Jord{\'a}n} {et~al.}(2020){Jord{\'a}n}, {Brahm}, {Espinoza},
  {Henning}, {Jones}, {Kossakowski}, {Sarkis}, {Trifonov}, {Rojas}, {Torres},
  {Drass}, {Nandakumar}, \& {Barbieri}}]{jordan:2020}
{Jord{\'a}n}, A., {Brahm}, R., {Espinoza}, N., {et~al.} 2020, \aj, 159, 145,
  \dodoi{10.3847/1538-3881/ab6f67}

\bibitem[{{Kaufer} {et~al.}(1999){Kaufer}, {Stahl}, {Tubbesing},
  {N{\o}rregaard}, {Avila}, {Francois}, {Pasquini}, \& {Pizzella}}]{kaufer:99}
{Kaufer}, A., {Stahl}, O., {Tubbesing}, S., {et~al.} 1999, The Messenger, 95, 8

\bibitem[{{Kipping} {et~al.}(2019){Kipping}, {Nesvorn{\'y}}, {Hartman},
  {Torres}, {Bakos}, {Jansen}, \& {Teachey}}]{kipp19}
{Kipping}, D., {Nesvorn{\'y}}, D., {Hartman}, J., {et~al.} 2019, \mnras, 486,
  4980, \dodoi{10.1093/mnras/stz1141}

\bibitem[{{Kipping}(2013{\natexlab{a}})}]{kipp13}
{Kipping}, D.~M. 2013{\natexlab{a}}, \mnras, 435, 2152,
  \dodoi{10.1093/mnras/stt1435}

\bibitem[{{Kipping}(2013{\natexlab{b}})}]{exoplanet:kipping13}
---. 2013{\natexlab{b}}, \mnras, 435, 2152, \dodoi{10.1093/mnras/stt1435}

\bibitem[{Korth(2020)}]{Korth2020}
Korth, J. 2020, PhD thesis.
\newblock \url{http://www.uni-koeln.de/}

\bibitem[{{K{\"u}rster} {et~al.}(2015){K{\"u}rster}, {Trifonov}, {Reffert},
  {Kostogryz}, \& {Rodler}}]{kurs15}
{K{\"u}rster}, M., {Trifonov}, T., {Reffert}, S., {Kostogryz}, N.~M., \&
  {Rodler}, F. 2015, \aap, 577, A103, \dodoi{10.1051/0004-6361/201525872}

\bibitem[{{Lee} \& {Chiang}(2016)}]{lee16}
{Lee}, E.~J., \& {Chiang}, E. 2016, \apj, 817, 90,
  \dodoi{10.3847/0004-637X/817/2/90}

\bibitem[{{Lee} \& {Peale}(2002)}]{lee02}
{Lee}, M.~H., \& {Peale}, S.~J. 2002, \apj, 567, 596, \dodoi{10.1086/338504}

\bibitem[{{Li} {et~al.}(2019){Li}, {Tenenbaum}, {Twicken}, {Burke}, {Jenkins},
  {Quintana}, {Rowe}, \& {Seader}}]{li19}
{Li}, J., {Tenenbaum}, P., {Twicken}, J.~D., {et~al.} 2019, \pasp, 131, 024506,
  \dodoi{10.1088/1538-3873/aaf44d}

\bibitem[{{Lithwick} \& {Naoz}(2011)}]{lith11}
{Lithwick}, Y., \& {Naoz}, S. 2011, \apj, 742, 94,
  \dodoi{10.1088/0004-637X/742/2/94}

\bibitem[{{Lithwick} {et~al.}(2012){Lithwick}, {Xie}, \& {Wu}}]{lith12}
{Lithwick}, Y., {Xie}, J., \& {Wu}, Y. 2012, \apj, 761, 122,
  \dodoi{10.1088/0004-637X/761/2/122}

\bibitem[{{Luger} {et~al.}(2019){Luger}, {Agol}, {Foreman-Mackey}, {Fleming},
  {Lustig-Yaeger}, \& {Deitrick}}]{exoplanet:luger18}
{Luger}, R., {Agol}, E., {Foreman-Mackey}, D., {et~al.} 2019, \aj, 157, 64,
  \dodoi{10.3847/1538-3881/aae8e5}

\bibitem[{{MacDonald} \& {Dawson}(2018)}]{macd18}
{MacDonald}, M.~G., \& {Dawson}, R.~I. 2018, \aj, 156, 228,
  \dodoi{10.3847/1538-3881/aae266}

\bibitem[{{Mandel} \& {Agol}(2002)}]{mand02}
{Mandel}, K., \& {Agol}, E. 2002, \apjl, 580, L171, \dodoi{10.1086/345520}

\bibitem[{{Markwardt}(2009)}]{mark09}
{Markwardt}, C.~B. 2009, in Astronomical Society of the Pacific Conference
  Series, Vol. 411, Astronomical Data Analysis Software and Systems XVIII, ed.
  D.~A. {Bohlender}, D.~{Durand}, \& P.~{Dowler}, 251.
\newblock \doarXiv{0902.2850}

\bibitem[{{Mayor} {et~al.}(2003){Mayor}, {Pepe}, {Queloz}, {Bouchy},
  {Rupprecht}, {Lo Curto}, {Avila}, {Benz}, {Bertaux}, {Bonfils}, {Dall},
  {Dekker}, {Delabre}, {Eckert}, {Fleury}, {Gilliotte}, {Gojak}, {Guzman},
  {Kohler}, {Lizon}, {Longinotti}, {Lovis}, {Megevand}, {Pasquini}, {Reyes},
  {Sivan}, {Sosnowska}, {Soto}, {Udry}, {van Kesteren}, {Weber}, \&
  {Weilenmann}}]{harps}
{Mayor}, M., {Pepe}, F., {Queloz}, D., {et~al.} 2003, The Messenger, 114, 20

\bibitem[{{McCully} {et~al.}(2018){McCully}, {Volgenau}, {Harbeck}, {Lister},
  {Saunders}, {Turner}, {Siiverd}, \& {Bowman}}]{McCully:2018}
{McCully}, C., {Volgenau}, N.~H., {Harbeck}, D.-R., {et~al.} 2018, in Society
  of Photo-Optical Instrumentation Engineers (SPIE) Conference Series, Vol.
  10707, \procspie, 107070K, \dodoi{10.1117/12.2314340}

\bibitem[{{McQuillan} {et~al.}(2014){McQuillan}, {Mazeh}, \&
  {Aigrain}}]{mcqu14}
{McQuillan}, A., {Mazeh}, T., \& {Aigrain}, S. 2014, \apjs, 211, 24,
  \dodoi{10.1088/0067-0049/211/2/24}

\bibitem[{{Millholland} {et~al.}(2018){Millholland}, {Laughlin}, {Teske},
  {Butler}, {Burt}, {Holden}, {Vogt}, {Crane}, {Shectman}, \&
  {Thompson}}]{mill18}
{Millholland}, S., {Laughlin}, G., {Teske}, J., {et~al.} 2018, \aj, 155, 106,
  \dodoi{10.3847/1538-3881/aaa894}

\bibitem[{{Mustill} {et~al.}(2015){Mustill}, {Davies}, \& {Johansen}}]{must15}
{Mustill}, A.~J., {Davies}, M.~B., \& {Johansen}, A. 2015, \apj, 808, 14,
  \dodoi{10.1088/0004-637X/808/1/14}

\bibitem[{{Nesvorn{\'y}} \& {Vokrouhlick{\'y}}(2016)}]{nesv16}
{Nesvorn{\'y}}, D., \& {Vokrouhlick{\'y}}, D. 2016, \apj, 823, 72,
  \dodoi{10.3847/0004-637X/823/2/72}

\bibitem[{{Parviainen}(2015)}]{parv15}
{Parviainen}, H. 2015, \mnras, 450, 3233, \dodoi{10.1093/mnras/stv894}

\bibitem[{{Rein} \& {Liu}(2012)}]{rein12}
{Rein}, H., \& {Liu}, S.~F. 2012, \aap, 537, A128,
  \dodoi{10.1051/0004-6361/201118085}

\bibitem[{{Rein} \& {Spiegel}(2015)}]{rein15}
{Rein}, H., \& {Spiegel}, D.~S. 2015, \mnras, 446, 1424,
  \dodoi{10.1093/mnras/stu2164}

\bibitem[{Salvatier {et~al.}(2016)Salvatier, Wiecki, \&
  Fonnesbeck}]{exoplanet:pymc3}
Salvatier, J., Wiecki, T.~V., \& Fonnesbeck, C. 2016, PeerJ Computer Science,
  2, e55

\bibitem[{{Schneider} {et~al.}(2011){Schneider}, {Dedieu}, {Le Sidaner},
  {Savalle}, \& {Zolotukhin}}]{schn11}
{Schneider}, J., {Dedieu}, C., {Le Sidaner}, P., {Savalle}, R., \&
  {Zolotukhin}, I. 2011, \aap, 532, A79, \dodoi{10.1051/0004-6361/201116713}

\bibitem[{{Smith} {et~al.}(2012){Smith}, {Stumpe}, {Van Cleve}, {Jenkins},
  {Barclay}, {Fanelli}, {Girouard}, {Kolodziejczak}, {McCauliff}, {Morris}, \&
  {Twicken}}]{smit12}
{Smith}, J.~C., {Stumpe}, M.~C., {Van Cleve}, J.~E., {et~al.} 2012, \pasp, 124,
  1000, \dodoi{10.1086/667697}

\bibitem[{{Stumpe} {et~al.}(2014){Stumpe}, {Smith}, {Catanzarite}, {Van Cleve},
  {Jenkins}, {Twicken}, \& {Girouard}}]{stum14}
{Stumpe}, M.~C., {Smith}, J.~C., {Catanzarite}, J.~H., {et~al.} 2014, \pasp,
  126, 100, \dodoi{10.1086/674989}

\bibitem[{{Tamayo} {et~al.}(2020){Tamayo}, {Rein}, {Shi}, \& {Hernand
  ez}}]{tama20}
{Tamayo}, D., {Rein}, H., {Shi}, P., \& {Hernand ez}, D.~M. 2020, \mnras, 491,
  2885, \dodoi{10.1093/mnras/stz2870}

\bibitem[{{Theano Development Team}(2016)}]{exoplanet:theano}
{Theano Development Team}. 2016, arXiv e-prints, abs/1605.02688.
\newblock \url{http://arxiv.org/abs/1605.02688}

\bibitem[{{Thommes} \& {Lissauer}(2003)}]{tho03}
{Thommes}, E.~W., \& {Lissauer}, J.~J. 2003, \apj, 597, 566,
  \dodoi{10.1086/378317}

\bibitem[{{Twicken} {et~al.}(2018){Twicken}, {Catanzarite}, {Clarke},
  {Girouard}, {Jenkins}, {Klaus}, {Li}, {McCauliff}, {Seader}, {Tenenbaum},
  {Wohler}, {Bryson}, {Burke}, {Caldwell}, {Haas}, {Henze}, \&
  {Sanderfer}}]{twic18}
{Twicken}, J.~D., {Catanzarite}, J.~H., {Clarke}, B.~D., {et~al.} 2018, \pasp,
  130, 064502, \dodoi{10.1088/1538-3873/aab694}

\bibitem[{{Van Eylen} {et~al.}(2019){Van Eylen}, {Albrecht}, {Huang},
  {MacDonald}, {Dawson}, {Cai}, {Foreman-Mackey}, {Lundkvist}, {Silva Aguirre},
  {Snellen}, \& {Winn}}]{vane19}
{Van Eylen}, V., {Albrecht}, S., {Huang}, X., {et~al.} 2019, \aj, 157, 61,
  \dodoi{10.3847/1538-3881/aaf22f}

\bibitem[{{Wolff} {et~al.}(2012){Wolff}, {Dawson}, \& {Murray-Clay}}]{wolf12}
{Wolff}, S., {Dawson}, R.~I., \& {Murray-Clay}, R.~A. 2012, \apj, 746, 171,
  \dodoi{10.1088/0004-637X/746/2/171}

\bibitem[{{Zechmeister} \& {K{\"u}rster}(2009)}]{zech09}
{Zechmeister}, M., \& {K{\"u}rster}, M. 2009, \aap, 496, 577,
  \dodoi{10.1051/0004-6361:200811296}

\end{thebibliography}

\clearpage

\begin{longtable}{lccccrl}
\caption{\label{tab:rvs} HARPS,FEROS, and PFS radial velocity measurements of TOI-216.}\\
\hline
\noalign{\smallskip}
$\rm BJD$ & RV  & $\sigma_\mathrm{RV}$ &    BIS  & $\sigma_\mathrm{BIS}$   & Inst. & Note\\
-2450000          & (m s$^{-1}$)  & (m s$^{-1}$)       &  (m s$^{-1}$) & (m s$^{-1}$)  &    \\
\hline
\endfirsthead
\caption{Continued.} \\
\hline
$\rm BJD$ & RV  & $\sigma_\mathrm{RV}$ &    BIS  & $\sigma_\mathrm{BIS}$   & Inst. & Note\\
-2450000          & (m s$^{-1}$)  & (m s$^{-1}$)       &  (m s$^{-1}$) & (m s$^{-1}$)  &    \\
\hline
\endhead
\hline
\endfoot
\noalign{\smallskip}
\noalign{\smallskip}
8449.70722 & 36690.0 &  7.8 & -16.0 & 12.0 & FEROS \\
8450.62181 & 36709.1 &  8.0 & -12.0 & 12.0 & FEROS \\
8450.75019 & 36686.8 &  8.3 & -35.0 & 12.0 & FEROS \\
8451.72708 & 36706.3 &  7.7 & -21.0 & 12.0 & FEROS \\
8451.74212 & 36696.3 &  7.2 & -44.0 & 11.0 & FEROS \\
8451.75750 & 36694.2 &  7.2 & -25.0 & 11.0 & FEROS \\
8452.77587 & 36713.0 &  7.6 & -48.0 & 11.0 & FEROS \\
8464.76381 & 36754.4 &  8.7 &  11.0 & 11.0 & HARPS \\
8464.81000 & 36744.7 & 10.3 &  19.0 & 13.0 & HARPS \\
8465.73524 & 36759.3 &  4.1 &   9.0 &  5.0 & HARPS \\
8465.75656 & 36753.4 &  4.6 &   5.0 &  6.0 & HARPS \\
8466.72641 & 36733.7 &  6.4 &  -5.0 &  8.0 & HARPS \\
8466.74802 & 36730.1 &  7.4 & -11.0 & 10.0 & HARPS \\
8467.74159 & 36714.5 &  7.9 & -35.0 & 12.0 & FEROS \\
8468.67914 &  10.49 & 1.11 & \dotfill & \dotfill & PFS \\
8468.74406 & 36727.7 &  7.7 &   0.0 & 12.0 & FEROS \\
8472.76399 &   1.38 & 1.23 & \dotfill & \dotfill & PFS \\
8476.74101 & -23.94 & 1.18 & \dotfill & \dotfill & PFS \\
8479.63075 & -33.40 & 1.15 & \dotfill & \dotfill & PFS \\
8480.71817 & -33.50 & 1.33 & \dotfill & \dotfill & PFS \\
8481.63918 & 36681.1 &  5.6 &   3.0 &  7.0 & HARPS \\
8481.66088 & 36670.1 &  4.9 &  -7.0 &  6.0 & HARPS \\
8482.63750 & 36690.5 &  4.1 &   2.0 &  5.0 & HARPS \\
8482.65892 & 36684.0 &  4.3 &  -1.0 &  6.0 & HARPS \\
8483.63626 & 36702.2 &  4.3 &  -8.0 &  6.0 & HARPS \\
8483.65821 & 36704.1 &  4.1 &  11.0 &  5.0 & HARPS \\
8483.76291 & 36668.1 &  8.0 &  -1.0 & 12.0 & FEROS \\
8485.70826 & 36650.9 & 13.9 &  47.0 & 19.0 & FEROS \\
8493.72558 & 36735.9 &  9.0 & -11.0 & 13.0 & FEROS \\
8497.62083 & 36696.5 &  7.8 & -40.0 & 12.0 & FEROS \\
8501.68699 &  34.11 & 1.72 & \dotfill & \dotfill & PFS \\
8503.65356 & 36813.1 & 14.3 &  64.0 & 19.0 & FEROS & Outlier \\
8504.65888 & 36684.4 &  9.0 & -61.0 & 13.0 & FEROS \\
8507.66569 & -11.68 & 2.42 & \dotfill & \dotfill & PFS \\
8510.68697 & -27.18 & 1.88 & \dotfill & \dotfill & PFS \\
8521.56775 & 36625.2 &  7.6 &   5.0 & 11.0 & FEROS & Outlier \\
8528.60948 &  46.62 & 1.76 & \dotfill & \dotfill & PFS \\
8529.58589 &  45.56 & 1.65 & \dotfill & \dotfill & PFS \\
8542.55798 & 36659.1 &  8.7 & -44.0 & 13.0 & FEROS \\
8543.54347 & 36640.3 &  7.4 &  -6.0 & 11.0 & FEROS \\
8544.61674 & 36663.3 &  7.8 & -26.0 & 12.0 & FEROS \\
8545.57452 & 36676.3 &  7.6 &  -3.0 & 11.0 & FEROS \\
8546.58251 & 36714.8 &  9.7 &  38.0 & 14.0 & FEROS \\
8547.53676 & 36635.1 &  7.9 & -17.0 & 12.0 & FEROS \\
8548.58309 & 36652.2 &  7.7 &  23.0 & 12.0 & FEROS \\
8550.61335 & 36613.1 &  8.2 & -62.0 & 12.0 & FEROS \\
8551.62488 & 36637.6 & 11.4 & -26.0 & 16.0 & FEROS \\
8554.56409 & 36707.0 &  8.9 & -38.0 & 13.0 & FEROS \\
8557.56708 & 36713.0 &  9.2 &  -4.0 & 13.0 & FEROS \\
8708.90979 &  28.50 & 2.14 & \dotfill & \dotfill & PFS \\
8711.89325 &  12.38 & 2.62 & \dotfill & \dotfill & PFS \\
8714.92414 & -16.98 & 2.14 & \dotfill & \dotfill & PFS \\
8742.88505 &  19.06 & 2.19 & \dotfill & \dotfill & PFS \\
8764.84159 & 36736.7 &  5.2 &  -1.0 &  7.0 & HARPS \\
8766.79176 & 36750.6 &  6.4 &   3.0 &  8.0 & HARPS \\
8767.86802 &  29.71 & 1.98 & \dotfill & \dotfill & PFS \\
8777.79725 & 36695.8 & 23.5 &  23.0 & 31.0 & HARPS \\
8914.55281 &  21.41 & 2.73 & \dotfill & \dotfill & PFS \\
8920.52884 &  -5.82 & 1.77 & \dotfill & \dotfill & PFS \\
8923.53817 & -33.84 & 1.94 & \dotfill & \dotfill & PFS \\
%\noalign{\smallskip}
\hline
%\noalign{\smallskip}
\end{longtable}
\clearpage
\begin{deluxetable}{rrl}
 \tablecaption{Example Fit with High Precision Parameters (Osculating Orbital Elements at Epoch 1325.3279 days) for \thisstarinn and \thisstarout Derived from joint TTV/RV fit \label{tab:ex} }
 \tablehead{
 \colhead{Parameter}    & \colhead{Value}}
 \startdata
 $M_\star (M_\odot)$ & 0.77\\
$M_\inn$ ($M_{\rm Jup}$) & 0.061381\\
$P_\inn$ & 17.096824\\
$e_\inn$ &0.15830\\
$\varpi_\inn$ (deg.) & 292.55926\\
$\lambda_\inn$ (deg) & 82.329609\\
$\Omega_{\inn, \rm sky}$ (deg) & 0 \\
$i_{\inn,\rm sky}$ (deg) & 88.594066\\
$M_\out$ ($M_{\rm Jup}$) &  0.55454\\
$P_\out$ &  34.551590\\
$e_\out$ &0.0056764\\
$\varpi_\out$ (deg.) & 200.93260\\
$\lambda_\out$ (deg) &  27.931018\\
$\Delta \Omega_{\rm sky} = \Omega_{\out, \rm sky} $ (deg) & -0.20544800\\
$i_{\inn,\rm sky}$ (deg) &  89.785658\\
\enddata
\end{deluxetable}

\end{document}